\documentclass[twocolumn]{aastex631}

\usepackage{subfigure}
\usepackage{amsmath}
\usepackage{mathrsfs}
\usepackage{esvect}

\received{March 1, 2024}
\revised{May 13, 2024}
\accepted{May 16, 2024}

\DeclareRobustCommand{\VAN}[3]{#2}
\let\VANthebibliography\thebibliography
\def\thebibliography{\DeclareRobustCommand{\VAN}[3]{##3}\VANthebibliography}

\begin{document}
\title{UVCANDELS: The role of dust on the stellar mass--size relation of disk galaxies at 0.5 $\leq$~$\boldsymbol{z}$~$\leq$ 3.0}

\correspondingauthor{Kalina~V.~Nedkova}
\email{knedkov1@jhu.edu}

\author[0000-0001-5294-8002]{Kalina~V.~Nedkova}
\affiliation{Department of Physics and Astronomy, Johns Hopkins University, 3400 North Charles Street, Baltimore, MD 21218, USA}
\affiliation{Space Telescope Science Institute, 3700 San Martin Drive, Baltimore, MD 21218, USA}

\author[0000-0002-9946-4731]{Marc Rafelski}
\affiliation{Space Telescope Science Institute, 3700 San Martin Drive, Baltimore, MD 21218, USA}
\affiliation{Department of Physics and Astronomy, Johns Hopkins University, 3400 North Charles Street, Baltimore, MD 21218, USA}

\author[0000-0002-7064-5424]{Harry~I.~Teplitz}
\affiliation{IPAC, Mail Code 314-6, California Institute of Technology, 1200 E. California Blvd., Pasadena CA, 91125, USA}

\author[0000-0001-7166-6035]{Vihang Mehta}
\affiliation{IPAC, Mail Code 314-6, California Institute of Technology, 1200 E. California Blvd., Pasadena CA, 91125, USA}

\author[0000-0001-9022-665X]{Laura DeGroot}
\affiliation{Department of Physics, The College of Wooster, 1189 Beall Avenue, Wooster, OH 44691, USA}

\author[0000-0002-5269-6527]{Swara Ravindranath}
\affiliation{Astrophysics Science Division, NASA Goddard Space Flight Center, 8800 Greenbelt Rd., Greenbelt, MD, 20771, USA}
\affiliation{Center for Research and Exploration in Space Science and Technology II, Department of Physics, Catholic University of America, 620 Michigan Ave N.E., Washington DC 20064, USA}

\author[0000-0002-8630-6435]{Anahita Alavi}
\affiliation{IPAC, Mail Code 314-6, California Institute of Technology, 1200 E. California Blvd., Pasadena CA, 91125, USA}

\author[0000-0001-7396-3578]{Alexander Beckett}
\affiliation{Space Telescope Science Institute, 3700 San Martin Drive, Baltimore, MD 21218, USA}


\author[0000-0001-9440-8872]{Norman~A.~Grogin}
\affiliation{Space Telescope Science Institute, 3700 San Martin Drive, Baltimore, MD 21218, USA}

\author[0000-0002-1857-2088]{Boris Häußler}
\affiliation{European Southern Observatory, Alonso de Cordova 3107, Casilla 19001, Santiago, Chile}

\author[0000-0002-6610-2048]{Anton~M.~Koekemoer}
\affiliation{Space Telescope Science Institute, 3700 San Martin Drive, Baltimore, MD 21218, USA}

\author[0000-0003-0028-4130]{Grecco A.~Oyarzún}
\affiliation{Department of Physics and Astronomy, Johns Hopkins University, 3400 North Charles Street, Baltimore, MD 21218, USA}

\author[0000-0002-0604-654X]{Laura Prichard}
\affiliation{Space Telescope Science Institute, 3700 San Martin Drive, Baltimore, MD 21218, USA}

\author[0000-0002-4917-7873]{Mitchell Revalski}
\affiliation{Space Telescope Science Institute, 3700 San Martin Drive, Baltimore, MD 21218, USA}


\author[0000-0002-4226-304X]{Gregory F.~Snyder}
\affiliation{Space Telescope Science Institute, 3700 San Martin Drive, Baltimore, MD 21218, USA}

\author[0000-0003-3759-8707]{Ben Sunnquist}
\affiliation{Space Telescope Science Institute, 3700 San Martin Drive, Baltimore, MD 21218, USA}

\author[0000-0002-9373-3865]{Xin Wang}
\affil{School of Astronomy and Space Science, University of Chinese Academy of Sciences (UCAS), Beijing 100049, China}
\affil{National Astronomical Observatories, Chinese Academy of Sciences, Beijing 100101, China}
\affil{Institute for Frontiers in Astronomy and Astrophysics, Beijing Normal University,  Beijing 102206, China}

\author[0000-0001-8156-6281]{Rogier A.~Windhorst}
\affiliation{School of Earth \& Space Exploration, Arizona State University, Tempe, AZ 85287-1404, USA}

\author[0000-0003-3691-937X]{Nima Chartab}
\affiliation{The Observatories of the Carnegie Institution for Science, 813 Santa Barbara St., Pasadena, CA 91101, USA}

\author[0000-0003-1949-7638]{Christopher J.~Conselice}
\affiliation{Jodrell Bank Centre for Astrophysics, University of Manchester, Oxford Road, Manchester M13 9PL, UK}

\author[0000-0003-2775-2002]{Yicheng Guo}
\affiliation{Department of Physics and Astronomy, University of Missouri, Columbia, MO 65211, USA}  

\author[0000-0001-6145-5090]{Nimish Hathi}
\affiliation{Space Telescope Science Institute, 3700 San Martin Drive, Baltimore, MD 21218, USA}                 

\author[0000-0001-8587-218X]{Matthew J.~Hayes}
\affiliation{Stockholm University, Department of Astronomy and Oskar Klein Centre for Cosmoparticle Physics, AlbaNova University Centre, SE-10691, Stockholm, Sweden}

\author[0000-0001-7673-2257]{Zhiyuan Ji}
\affiliation{Steward Observatory, University of Arizona, 933 N. Cherry Avenue, Tucson, AZ 85721, USA}

\author[0000-0001-6505-0293]{Keunho J.~Kim}
\affiliation{IPAC, Mail Code 314-6, California Institute of Technology, 1200 E. California Blvd., Pasadena CA, 91125, USA}

\author[0000-0003-1581-7825]{Ray A.~Lucas}
\affiliation{Space Telescope Science Institute, 3700 San Martin Drive, Baltimore, MD 21218, USA}

\author[0000-0001-5846-4404]{Bahram Mobasher}
\affiliation{Department of Physics and Astronomy, University of California, Riverside, 900 University Ave, Riverside, CA 92521, USA}

\author[0000-0002-8190-7573]{Robert W.~O'Connell}
\affiliation{Department of Astronomy, University of Virginia, Charlottesville, VA 22904-4325, USA}

\author[0000-0002-0364-1159]{Zahra Sattari}
\affiliation{Department of Physics and Astronomy, University of California, Riverside, 900 University Ave, Riverside, CA 92521, USA}
\affiliation{The Observatories of the Carnegie Institution for Science, 813 Santa Barbara St., Pasadena, CA 91101, USA}

\author[0000-0002-0648-1699]{Brent M.~Smith}
\affiliation{School of Earth \& Space Exploration, Arizona State University, Tempe, AZ 85287-1404, USA}

\author[0000-0003-0749-4667]{Sina Taamoli}
\affiliation{Department of Physics and Astronomy, University of California, Riverside, 900 University Ave, Riverside, CA 92521, USA}

\author[0000-0003-3466-035X]{{L.~Y.~Aaron} {Yung}}
\affiliation{Space Telescope Science Institute, 3700 San Martin Drive, Baltimore, MD 21218, USA}

\author{the UVCANDELS Team}





\begin{abstract}
We use the Ultraviolet Imaging of the Cosmic Assembly Near-infrared Deep Extragalactic Legacy Survey fields (UVCANDELS) to measure half-light radii in the rest-frame far-UV for $\sim$16,000 disk-like galaxies over $0.5\leq z \leq 3$. We compare these results to rest-frame optical sizes that we measure in a self-consistent way and find that the stellar mass--size relation of disk galaxies is steeper in the rest-frame UV than in the optical across our entire redshift range. We show that this is mainly driven by massive galaxies ($\gtrsim10^{10}$M$_\odot$), which we find to also be among the most dusty. Our results are consistent with the literature and have commonly been interpreted as evidence of inside-out growth wherein galaxies form their central structures first. However, they could also suggest that the centers of massive galaxies are more heavily attenuated than their outskirts. We distinguish between these scenarios by modeling and selecting galaxies at $z=2$ from the VELA simulation suite in a way that is consistent with UVCANDELS. We show that the effects of dust alone can account for the size differences we measure at $z=2$. This indicates that, at different wavelengths, size differences and the different slopes of the stellar mass--size relation do not constitute evidence for inside-out growth.

\end{abstract}


\section{Introduction}\label{sec:intro}
Galaxy size is a key observable property that 
offers insights into the evolutionary mechanisms that shape galaxies.
Despite the complex and nonlinear nature of these evolutionary mechanisms, galaxies have been observed to lie on remarkably tight relations on the size--stellar mass plane, known as the stellar mass--size relations. These relations describe the underlying spatial distribution of stellar mass in galaxies and have been extensively studied in an effort to probe how galaxies have built up their mass over cosmic time \citep[e.g.][]{Mancini2010MNRAS, vdWel2014, Mowla2019ApJ, Dimauro2019, Kalina2021, Kawinwanichakij2021ApJ, Cutler2022ApJ...925...34C, Ward2024ApJ, Buitrago2024A&A}. 
Three main tenets of galaxy size evolution have been established over the last couple of decades. First, more massive galaxies are on average larger than their lower mass counterparts. Second, galaxies were smaller at higher redshift, and third, at high stellar masses (i.e.~M$_\star \gtrsim 10^9$M$_\odot$), star-forming galaxies lie on a shallower stellar mass--size relation and are typically more extended than quiescent galaxies.

The interpretation of these results has been a source of some debate in recent years. One interpretation is that galaxies evolve on the mass-size plane by forming new stars mostly within their disk structures until they reach a central density threshold, at which they begin to quench \citep[e.g.][]{vdwel2009ApJ, Williams2010ApJW, Dokkum2015ApJ, Chen2020ApJ...897..102C}. Once these galaxies quench, their growth is dominated by dry minor mergers, which typically leave galaxy light profile shapes intact but result in significant size evolution \citep[e.g.][]{Buitrago2008ApJ,Bezanson2009ApJ, Naab2009ApJ, Oser2012ApJ, Bluck2012ApJ, Ji2022ApJ} and variations in their chemical composition \citep{Cook2016ApJ, Grecco2019ApJ}. This picture successfully explains the observed trends in the stellar mass--size relations of both massive star-forming and quiescent galaxies.

More recently, however, \cite{Mosleh2017ApJ...837....2M}, \cite{ Suess2019ApJ, Suess2019ApJ...885L}, \cite{ Miller2023ApJ...945..155M}, and others, have argued that stellar mass--size relations should be derived using half-mass radii, which result in significantly shallower stellar mass--size relations. As radial color gradients can bias light-weighted sizes (see also \citealt{Buitrago2024A&A} and \citealt{Chamba2023arXiv}), these studies have suggested that the observed redshift evolution in the stellar mass--size relation, when half-light radii are used, is primarily caused by these radial color gradients.
Luckily, software suites with multi-wavelength fitting capabilities, like the Mega-Morph tools \citep{MegaMorph, Vika2013MNRAS, MegaMorph2} that we use in this work, are well suited to address this issue. Indeed, \cite{vdwel2024ApJ} who use these same tools to measure galaxies' physical properties, show that similar characteristics and trends are present in the stellar mass--size relation regardless of whether light- or mass-weighted sizes are used. This indicates that studying the evolution of half-light radii can reliably inform our understanding of how galaxies build up their stellar mass.

Still, an important consideration when interpreting stellar mass--size relations of galaxies is that sizes vary as a function of wavelength  \citep[e.g.][]{Evans1994MNRAS, LaBarbera2002ApJ, LaBarbera2010MNRAS, MegaMorph, Vulcani2014MNRAS, Kennedy2016MNRAS}. This is because different stellar populations within galaxies usually have distinct spatial distributions, with young, hot, blue stars typically residing in the outer disk structures while the bulk of galaxies' older stellar populations are often found towards their centers in the, typically smaller, bulges \citep[e.g.][]{deJong1996A&A, Peletier1996AJ,Bell2000MNRAS, MacArthur2004ApJS..152..175M, MunozMateos2007ApJ, MunozMateos2009ApJ, Sanches-Bazquez2014A&A, GonzalezDelgado2014A&A, Zheng2017MNRAS, Pessa2023A&A,Jegatheesan2024A&A}.
As blue and red wavelengths are sensitive to the young and old stellar populations within galaxies, respectively, most star-forming galaxies appear larger at blue wavelengths than at redder wavelengths \citep[e.g.][]{Kelvin2012MNRAS.421.1007K, Vulcani2014MNRAS, Kennedy2016MNRAS, Baes2024A&A}. 
Although this wavelength dependence poses a slight challenge in measuring galaxy sizes consistently, it also presents an opportunity to probe how galaxies build up their stellar populations by studying the stellar mass--size relations in different wavelength regimes.

Previous works have shown that massive galaxies are typically larger in the rest-frame UV than rest-frame optical \citep[e.g.][although see \citealt{Curtis-Lake2016MNRAS} who find similar sizes at UV and optical wavelengths]{Papovich2005ApJ,Shibuya2015, Mosleh2012, Cheng2020A&A...633A.105C, Ji2024arXiv, Angelo2024MNRAS}. The observed difference in size between the two wavelength regimes is expected for inside-out growth \citep{Dutton2011MNRAS}. In this model, galaxies first form their centers and then grow their outskirts via star formation at later times. This results in galaxies with older, redder stellar populations in their inner regions and younger, bluer stellar populations in their outer disks. Another physical explanation is the so-called `outshining effect' \citep{Maraston2010MNRAS, Reddy2012ApJ, Wuyts2012ApJ} where galaxies' light is dominated by their youngest stellar populations, which outshine the older ones.
This results in brighter UV
luminosities in galaxies’ outskirts leading to lower Sérsic indices and larger rest-frame UV sizes
than optical sizes.

However, dust also plays a significant role in our understanding of galaxies, complicating the interpretation of observations, and the derivation of fundamental properties such as galaxy stellar masses and star formation rates \citep[e.g.][]{Hopkins2001AJ....122..288H, Sullivan2001ApJ...558...72S, Taylor2011, Leja2019ApJ...877..140L}. Dust obscuration tends to reduce the emergent light from galaxies in UV to near-infrared wavelengths through scattering and absorption \citep[e.g.][]{Witt1992ApJ,Witt2000ApJ...528..799W}, the net effect of which is termed attenuation. Hence, the differences in UV and optical sizes measured in previous works could be explained by the outshining effect, inside-out growth, galaxies being more heavily attenuated in their centers, or a combination of these.

Cosmological hydrodynamic simulations provide a unique opportunity to gain insight into the contribution of each effect on galaxy sizes. For instance, \cite{Marshall2022MNRAS} used the \textsc{BlueTides} simulation suite \citep{Feng2015ApJ} to explore the size--luminosity relation in the rest-frame UV and optical. They find that the slope of the relations decreases at longer wavelengths and argue that this is a direct result of dust, which produces less attenuation at redder wavelengths. 
Additionally, from the \textsc{SIMBA} simulations \citep{Simba2019MNRAS}, \cite{Wu2020MNRAS} predict that the sizes of star-forming galaxies at the epoch of reionization are consistent in the UV and optical as they find that the centers of the galaxies in \textsc{SIMBA} are younger but also more dust attenuated. They suggest that combined, these two effects weaken the color gradients in the mock galaxies, resulting in similar sizes in the UV and optical. As can be seen, simulations are a powerful tool for understanding how dust affects observed properties of galaxies at different wavelengths, but a consensus has not yet been reached.

In this work, we set out to shed light on this question by measuring galaxy sizes in the rest-frame UV and optical using observations from 
the Ultraviolet Imaging of the Cosmic Assembly Near-infrared Deep Extragalactic Legacy Survey fields (UVCANDELS; PI: H.~I.~Teplitz) and simulations in a consistent way. UVCANDELS provides high-resolution UV imaging of galaxies over the largest area to date \citep{Wang2024RNAAS}. Galaxy properties in the UV are particularly important as the rest-frame UV probes ongoing star formation. As such, UVCANDELS has already been used to explore the properties of clumpy galaxies \citep[][]{Sattari2023ApJ,Martin2023ApJ}, the star formation burstiness of galaxies by comparing their UV and H$\alpha$ emission \citep{Mehta2023ApJ}, the environmental dependence of quenching processes \citep{Kuschel2023ApJ}, Lyman continuum emission from AGN \citep{Smith2024arXiv} and star-forming galaxies \citep{Wang2023arXiv230809064W}, the UV luminosity function \citep{Sun2023arXiv}, and the UV spectral slope of star-forming galaxies \citep{Morales2024arXiv}.
The UVCANDELS data also provide the unique opportunity to investigate the properties of galaxies in already very well studied fields with a wealth of ancillary data. Using these data, we measure rest-frame UV sizes at lower redshifts than previously possible and compare these with results from the VELA simulation suite \citep{Ceverino2014MNRAS} to yield new insights into the buildup of galaxies.


The layout of this paper is as follows. In \S\ref{sec:data}, we discuss the UVCANDELS data as well as the simulated galaxy sample from the VELA simulation suite. The technical aspects of modeling the galaxies and obtaining their rest-frame UV and optical sizes are also discussed in this section. In \S\ref{sec:results}, we present the rest-frame UV and optical stellar mass--size relations for disk galaxies in UVCANDELS and investigate how dust may be affecting these results. In \S\ref{sec:mre_from_sim}, we present the same results but from mock galaxies that are modeled and selected in a way that is consistent with the UVCANDELS data. We discuss the implications of our findings in \S \ref{sec:discussion} and provide a summary in \S\ref{sec:conclusions}. Throughout this paper, we use AB magnitudes \citep{Oke1983}. We also use cosmological density parameters $\Omega_{\mathrm{m}}$ = 0.27 and $\Omega_\Lambda$ = 0.73 and a Hubble constant of H$_{0} = 70$ km s$^{-1}$ Mpc$^{-1}$. In all figures, we adopt a convention of plotting results from UVCANDELS as circles and simulations as squares. 

\section{Data} \label{sec:data}

In this section, we first describe the UVCANDELS dataset and its properties in \S\ref{sec:uvcandels}. We then present the quality cuts applied to select reliable samples of disk galaxies in \S\ref{sec:selection}. As many of our selection criteria are motivated by the software suite that is used to model the data, the MegaMorph tools are discussed in this subsection as well. Finally, in \S\ref{sec:simulateddata}, we describe the mock galaxy sample from the VELA simulation suite.

\subsection{UVCANDELS} \label{sec:uvcandels}

\cite{Wang2024RNAAS} provide a comprehensive overview of the UVCANDELS Treasury Program as well as details about the observing strategy and data reduction. We include a brief summary of the aspects that are relevant for this work here. UVCANDELS provides imaging in the HST WFC3/UVIS F275W filter for four of the five CANDELS fields \citep{Grogin2011ApJS..197...35G, Koekemoer2011ApJS}: GOODS-N, GOODS-S, EGS, and COSMOS, as well as ACS/WFC F435W imaging in the COSMOS and EGS fields. This Cycle 26 HST program acquired 164 orbits of imaging, covering a total area of $\sim$430 arcmin$^2$, and reaches a $5\sigma$ depth of $m_{\mathrm{AB}} \approx$ 27 and 28 in F275W and F435W, respectively. 

The focus of this work is measuring the rest-frame UV sizes of galaxies at 1500{\AA} and comparing these to rest-frame optical sizes measured using CANDELS data at 5000{\AA}, the latter of which have already been studied for large galaxy samples across a wide range of redshifts \citep[e.g.][]{Shen2003MNRAS, Ferguson2004ApJ, Ravindranath2004ApJ,  Huang2013ApJ, Lange2016, Mowla2019ApJ, Kalina2021, Kawinwanichakij2021ApJ, Ward2024ApJ, Ji2024arXiv}. The F275W and F435W bands are essential for galaxies at $0.5\lesssim z\lesssim 1.3$ and $1.5\lesssim z\lesssim 3.0$, respectively, because they allow the rest-frame 1500{\AA} properties to be measured at these redshifts. We highlight here that most previous works that have explored the UV sizes of galaxies \citep[e.g.][]{Papovich2005ApJ, Shibuya2015, Mosleh2012, Curtis-Lake2016MNRAS} have been restricted to higher redshifts due to a lack of UV and blue-optical imaging. The imaging obtained as part of the UVCANDELS program therefore presents a crucial step forward in understanding how the UV sizes of galaxies evolve in largely unexplored redshift regimes. Table~\ref{tab:Bands_used} provides a list of the bands that we use for each field.

\begingroup
\setlength{\tabcolsep}{5pt} 
\renewcommand{\arraystretch}{1} 
\begin{table}
\caption{List of HST ACS and WFC3 filters used for each field, with the ones in bold indicating imaging provided by UVCANDELS \citep{Wang2024RNAAS}.}
\centering
\begin{tabular}{l l }
\hline
Fields &  Filters Used \\
\hline 
GOODS-N \& GOODS-S & \textbf{F275W}, F435W, F606W, \\
& F775W, F814W,  F850LP,  \\
& F105W,  F125W, F140W, \\&F160W \\
COSMOS \& EGS &  \textbf{F275W},  \textbf{F435W}, F606W,  \\
 & F814W, F125W, F140W,\\
 &F160W\\
\hline
\end{tabular}
\label{tab:Bands_used}
\end{table}
\endgroup

\subsubsection{UV Imaging}\label{sec:211}

Throughout this work, we use 30mas UVCANDELS images, publicly available on MAST\footnote{\url{https://archive.stsci.edu/hlsp/uvcandels}} combined with CANDELS data of the same pixel scale\footnote{\url{https://archive.stsci.edu/hlsp/candels}} \citep{Grogin2011ApJS..197...35G, Koekemoer2011ApJS}. For each of the four UVCANDELS fields, we create stacked white light images to use as our detection images. This is motivated by \cite{Lange2016} who find that the fitting results from \textsc{Galfit-3} \citep{Peng2010Galfit} can be significantly impacted by the initial starting parameters. These starting parameters are determined by running \textsc{SExtractor} \citep{Bertin1996} on a single detection image and are then used as the initial starting values for each filter in the multi-band fitting. Although \cite{MegaMorph} have demonstrated that using multi-wavelength data to model galaxy light profiles increases the accuracy of the fits, we create white light images to obtain more reliable starting parameters. 
Additionally, these white light images have higher signal-to-noise (SNR) than single-band detection images, enabling a reliable detection of even the faintest sources. They also allow differences in morphology at different bands to be accounted for.

\begin{figure*}
    \centering
    \includegraphics[width=1.\textwidth]{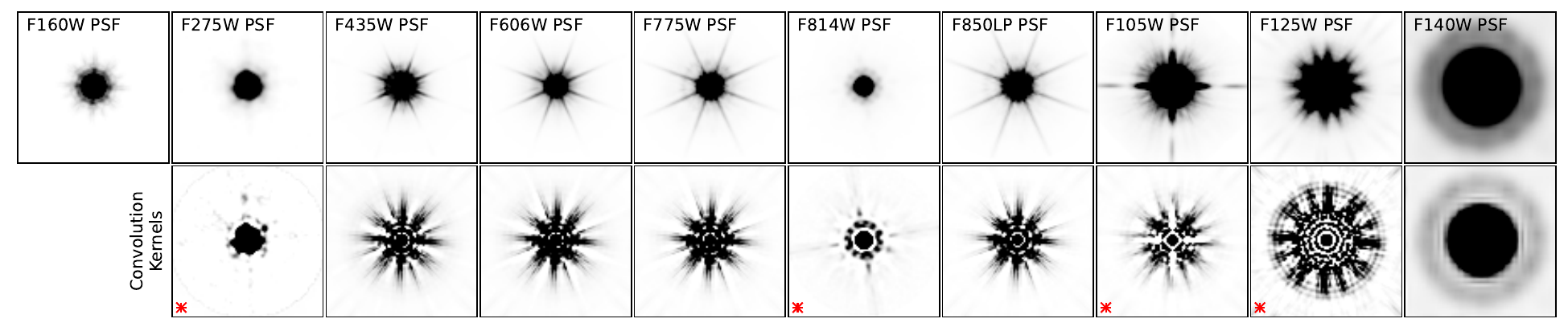}
    \caption{\textit{top}: Point spread functions (PSFs) for each band used in the GOODS-N field. \textit{bottom}: Convolution kernels used to match each PSF to the F160W PSF to create white light images for reliable object detection. We use Hanning window functions to create all matching convolution kernels, except those indicated with a red asterisk at the bottom left, for which we use cosine bell window functions where 30\% of the array values are tapered. See \S\ref{sec:211} for more details.} 
    \label{fig:PSFs+kernels}
\end{figure*}

To construct white light images, we first generate convolution kernels that we use to match the point spread function (PSF) of each image to the F160W PSF so the spatial resolution of each mosaic is consistent. We use F275W and F435W PSFs that were created from the UVCANDELS imaging by the IRAF DAOPHOT package \citep{Stetson1987PASP} using well-exposed, non-saturated stars as inputs. PSFs for the remaining filters (i.e.~those that are not in bold in Table~\ref{tab:Bands_used}) were created by the CANDELS collaboration. Using these PSFs, we generate convolution kernels following an approach similar to the one used in \cite{Revalski2023}. Specifically, we use the \href{https://photutils.readthedocs.io/en/stable/psf_matching.html}{PSF Matching} feature from the Astropy photutils package, and adopt Hanning window functions to create all convolution kernels, except those marked with a red asterisk in Figure~\ref{fig:PSFs+kernels}, for which we use cosine bell window functions where 30\% of the array values are tapered. Our choice of PSF convolution window function is based on visual inspection of the convolution kernels to reduce asymmetric features and nonphysical patterns.


The mosaics are then convolved with their respective kernels. In Figure \ref{fig:PSFs+kernels}, we show the PSFs for each of the GOODS-N filters that we use and the convolutions kernels used to match the spatial resolution of each mosaic to the F160W mosaic.
Lastly, the convolved images are matched to the same zero-point and stacked to produce the white light image. We note that we also create stacked root mean squared (RMS) images for each field in order to properly weight our detection image. These are created by adding the RMS map values at each pixel from all available filters in quadrature and dividing by the square root of the number of available bands.

\begingroup
\setlength{\tabcolsep}{2pt} 
\renewcommand{\arraystretch}{0.9} 
\begin{table}[]
    \centering
    \caption{Key \textsc{SExtractor} parameters used for the source detection on our white light images. We use the same parameters across all four UVCANDELS fields and provide the values used in our hot- and cold-mode runs.}
    \begin{tabular}{l c c}
        Parameter & Hot-Mode & Cold-Mode \\
        \hline
        \texttt{DETECT\_MINAREA} & 3.0 & 20.0\\
        \texttt{DETECT\_THRESH} & 1.1 & 5.0\\
        \texttt{ANALYSIS\_THRESH} & 1.1 & 5.0\\
        \texttt{FILTER\_NAME} & gauss\_4.0\_7x7.conv& tophat\_9.0\_9x9.conv\\
        \texttt{DEBLEND\_MINCONT} & 0.0001 & 0.01 \\
        \texttt{DEBLEND\_NTHRESH} & 64 & 32\\
        \hline
    \end{tabular}
    \label{tab:SExtractor_params}
\end{table}
\endgroup

To obtain starting parameters for our modeling, we run hot- and cold-mode \textsc{SExtractor} on the white light images following e.g. \cite{Barden2012} and \cite{Guo2013ApJS}. We run both modes to detect faint sources while also reducing the number galaxies that are falsely broken up into multiple objects (e.g.~due to internal structure). The \textsc{SExtractor} parameters that we use are provided in Table~\ref{tab:SExtractor_params} for repeatability, noting that the cold-mode runs are intended to detect only the most extended objects. We also visually inspect the source detection and use built-in features of the MegaMorph tools to flag any `false' detections or objects. These can be one of two categories. The first includes objects that are not real such as hot pixels along the edges or clumps that are part of a larger galaxy. These are removed before staring the modeling. The second category includes features such as diffraction spikes, which must be modeled so that they do not affect the fits of `nearby' objects.
These objects are removed from the final catalog, after all objects have been modeled.

\subsubsection{Photometric Redshifts and Stellar Masses}\label{sec:photo_z_M*}

Including UV imaging provides significant enhancements to the data products. In particular, photometric redshift estimates are significantly improved by reducing the number of catastrophic failures \citep[][Mehta et al.~\textit{in prep}]{Rafelski2015AJ} because imaging in the UV allows the Balmer break at low redshift to be distinguished from the Lyman break for higher redshift galaxies. Throughout this study, we use spectroscopic redshifts where they are available and photometric redshifts measured by the UVCANDELS team (Mehta et al.~\textit{in prep}) for the majority of our sample. In brief, these photometric redshifts are obtained from several different codes: EAZY \citep{eazy_brammer}, BPZ \citep{Coe2006AJ}, LePhare \citep{Arnouts1999MNRAS,Ilbert2006A&A}, and zphot \citep{Giallongo1998AJ, Fontana2000AJ....120.2206F}. The results from these are then combined largely following the procedure outlined in \cite{Dahlen2013ApJ}.


The derivation of stellar masses and other physical properties will also be outlined in Mehta et al.~(\textit{in prep}). To obtain these, spectral energy distributions (SEDs) are fit to multi-band photometry from the UV to infrared using \textsc{Dense Basis} \citep{Iyer2019ApJ}, assuming a \cite{Chabrier2003} initial mass function and \cite{Calzetti2000} dust law. The infrared photometry used is from the CANDELS photometric catalogs \citep{Guo2013ApJS, Nayyeri2017ApJS, Stefanon2017ApJS, Barro2019ApJS} which describe these data fully. Briefly, the \textit{Spitzer}/IRAC data for these CANDELS fields were obtained as part of the S-CANDELS$\footnote{https://www.ipac.caltech.edu/doi/irsa/10.26131/IRSA419}$ \citep{Ashby2015ApJS}, SEDS \citep{Ashby2013ApJ}, AEGIS \citep{Barmby2008ApJS}, S-COSMOS \citep{Sanders2007ApJS} and Spitzer GOODS \citep{Giavalisco2004ApJ} surveys. To fit the data across all available wavelengths, 
\textsc{Dense Basis} uses nonparametric descriptions of galaxy star formation histories (SFHs), which have been shown to produce more reliable results and uncertainties over parametric models, as they are flexible enough to describe the full diversity of SFH shapes (see \citealt{Pacifici2023ApJ} for detailed comparisons of commonly used SED fitting codes).

In order to obtain galaxy sizes, we fit the light profiles of galaxies with Sérsic profiles that are integrated to infinity to obtain the total magnitude. As this is different from the aperture-matched photometry on which the \textsc{Dense Basis} code is run to obtain stellar masses, we apply a correction to our stellar mass estimates following \cite{vdWel2014} and \cite{Kalina2021}. Specifically, we correct for the difference between the total flux from the photometric catalog and the total flux measured with the MegaMorph tools in the F160W filter.
We find that that this correction is consistent with zero except for the most massive galaxies at $\sim 10^{11.5}$M$_\odot$ for which it increases to $\sim$0.02 dex. Even at these highest masses, this correction is significantly smaller than our typical stellar mass uncertainties of $\sim$0.2 dex. We additionally find that the stellar mass correction does not show any strong dependence on galaxies' UV luminosity. Therefore, while we apply this correction to ensure that our size and mass estimates are based on the same model of the light distribution, we find that it does not affect the results presented in this paper. 

\subsection{Sample Selection}\label{sec:selection}




Many of our sample selection criteria are motivated by our fitting approach, therefore, we begin this section with a brief discussion of the MegaMorph ($\textbf{Me}$asuring $\textbf{Ga}$laxy $\textbf{Morph}$ology) tools. These tools consist of \textsc{GalfitM} and \textsc{Galapagos-2}, which build on \textsc{Galfit} \citep{Peng2002AJ, Peng2010Galfit} and \textsc{Galapagos} \citep{Barden2012}, respectively. They allow robust size measurements as well as a characterization of morphological properties to be reliably established by using and modeling data at all available wavelengths simultaneously.
\cite{MegaMorph, MegaMorph2} provide detailed discussions of the benefits of modeling galaxies with a multi-wavelength approach. We refer to those studies for more details about the software as well as for justifications regarding the functional forms with which each parameter is fit and the imposed limits, but briefly discuss key aspects that are relevant for this work below.  


Throughout this work, we model all UVCANDELS galaxies with a single Sérsic component such that the global properties of disk-like galaxies can be recovered. This modeling is roughly consistent with that carried out by \cite{Kalina2021} as we use the same software and CANDELS images but here, we also include additional F275W and F435W imaging, which is crucial for this study. We find that our size measurements are consistent with \cite{Kalina2021} in the wavelength ranges covered by both studies, suggesting that our measurements are robust. Given these similarities, most of the parameter constraints discussed below are motivated by that work.

We begin by briefly discussing the setup used in \textsc{Galapagos-2}. Recent results from \cite{Angelo2024MNRAS} indicate that half-light radius estimates can vary significantly depending on objects' cutout sizes. In light of this, we note that to create cutouts \textsc{Galapagos-2} uses Kron radii estimated with \textsc{SExtractor} and increases these by a user-specifiable factor. We have set this factor to three to be conservative. \cite{Angelo2024MNRAS} show that using a sufficiently large cutout size is vital because a crucial part of galaxy surface brightness profile fitting is obtaining precise background measurements as they can significantly impact other fit parameters. These background measurements will be biased if the image does not contain sufficient `empty' sky pixels \citep{Barden2012}. We emphasize that while \cite{Angelo2024MNRAS} estimate the sky background from their cutouts, \textsc{Galapagos-2} 
provides an estimate from the full mosaic \textit{before} the \textsc{GalfitM} fitting is started. In brief, the average flux is measured in elliptical annuli centered on each central object, where any flux from neighboring sources that fall within the annuli is excluded. This gives average flux as a function of radius away from the central object. Once the average flux begins to remain constant with radius, \textsc{Galapagos-2} determines the background from the last few annuli \citep[see][for more details]{Barden2012}. 
We find that this method works well as we have visually inspected $>$800 galaxy fits, finding no instances where the cutout size or sky background estimate hinder the galaxy parameter estimation.

\begin{figure}
    \centering
    \includegraphics[width=0.47\textwidth]{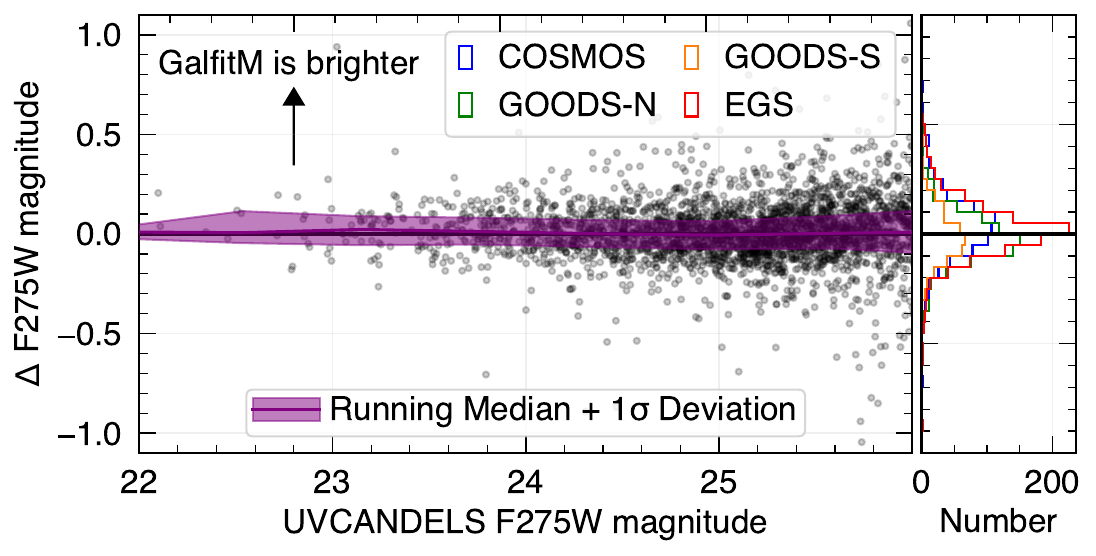}
    \caption{Comparison of the F275W magnitude from the UVCANDELS photometric catalogs \protect\citep{Sun2023arXiv} to the F275W magnitudes that we measure with \textsc{GalfitM} for galaxies at $0.5\leq z \leq1.3$, where we use the F275W band to constrain the rest-frame UV sizes of galaxies. In general, we find good agreement as indicated by the running median and $1\sigma$ deviation, shown in purple. The histogram shows the distribution of the magnitude difference for each field.}
    \label{fig:mag_comp}
\end{figure}

The remaining parameters in the \textsc{Galapagos-2} setup are matched to those used in \cite{Kalina2021}. These are discussed in more detail below, where we present the degrees of freedom with which each fit parameter is allowed to vary as a function of wavelength. Specifically, in our modeling, galaxies' magnitudes are allowed to vary freely as a function of wavelength, 
as is commonly adopted in the literature \citep[e.g.][]{MegaMorph, Kalina2021, Ward2024ApJ}. In Figure~\ref{fig:mag_comp}, we compare the F275W magnitudes of galaxies at $0.5\leq z\leq1.3$ from the UVCANDELS photometric catalogs \citep{Sun2023arXiv} to those that we derive with \textsc{GalfitM}. 
We find excellent agreement despite using different detection images. \cite{Sun2023arXiv} use 60mas F275W mosaics, while we generate white light images for each field from all available 30mas images, as discussed in \S\ref{sec:211}.
Similar to \cite{Nedkova2024arXiv}, we find that objects that are measured as being fainter with \textsc{GalfitM}
tend to be ‘near’ bright objects. Despite careful UV-optimized aperture photometry, which uses V-band isophotes as described in \cite{Sun2023arXiv}, the UVCANDELS photometric catalogs are typically impacted in such cases as it is challenging to entirely remove contamination from nearby bright neighbors when using apertures.
Conversely, objects that have significantly brighter \textsc{GalfitM}-derived magnitudes tend to have unreliable fits, for instance because the model is also trying to fit a neighbor.
We find that requiring the magnitude difference to be smaller than 0.5 removes the majority of these cases and therefore, we apply this as one of our selection criteria in \S\ref{sec:subselection}. This cut removes a total of 76 galaxies. We have visually inspected the fits in all bands for each object, finding clear reasons why each should be removed from our sample. In particular, we find that 63 objects have segmentation maps from our modeling that are inconsistent with those obtained by \cite{Sun2023arXiv} (e.g.~objects that are treated as a single system in one but split up into different objects in the other). The other 13 objects have fits where \textsc{GalfitM} has incorporated light from a neighboring object into the model and has thus returned unreliable parameter estimates. While these 76 objects should be excluded as their positions on the stellar mass--size relation will be wrong, we note that including them does not change the measured stellar mass--size relations that will be presented in the following sections. 

Although we allow galaxy magnitudes to vary freely as a function of wavelength, other parameters, such as galaxy sizes and Sérsic indices, are expected to vary smoothly, for instance, due to the presence of bulge and disk structures within disk-like galaxies \citep[e.g.][]{MegaMorph, Vulcani2014MNRAS, Kim2018ApJ, Nedkova2024arXiv}. We therefore fit the wavelength dependence of these parameters with second order polynomials. We note here that recent results from \cite{Ward2024ApJ}
indicate that for most galaxies, there exists at least one inflection point in the wavelength dependence of galaxy sizes and Sérsic indices. Thus, they suggest that a third order polynomial may be a more appropriate choice for their data, which includes infrared imaging from JWST. Within the wavelength range explored in this study, we find second order polynomials can sufficiently model these parameters, as they can accurately reproduce the data when one inflection point exists.

Lastly, a third set of parameters would be expected to remain roughly constant with wavelength, including the center coordinates of the profiles, axis ratios, and position angles. We choose to fit these with zeroth order polynomials such that they are free parameters in the fit but are held constant as a function of wavelength. We note here that allowing these parameters to instead vary smoothly with wavelength, in the same way that the half-light radii and Sérsic indices do, does not significantly alter the size measurements that \textsc{GalfitM} returns. Specifically, in comparing the half-light radii from two independent sets of models (one where the center coordinates, axis ratios, and position angles are allowed to vary smoothly with wavelength and a second where they are held constant), we measure an RMS of only 0.007 pixels with no systematic offsets. However, differences in the half-light radii measurements tend to be larger for galaxies that have bright blue star-forming clumps. For these galaxies, the location of the brightest clump is often identified as the central position in the bluest filters. In such cases, we find that the recovered half-light radius in the F275W band can differ by up to $\sim$20\% between the two models. As clumps can significantly influence the central positions measured in blue bands, we argue that keeping the center coordinates of the profiles constant with wavelength is a better choice. 
Overall, given the small RMS of the difference in effective radius, we find that keeping these parameters constant with wavelength is a reasonable approximation for our modeling, similar to \cite{MegaMorph}.


\subsubsection{Selecting Disk-Like Galaxies}\label{sec:subselection}

Due to the similarities between the modeling carried out in this work and by \cite{Kalina2021}, many of our selection criteria are also motivated by that work. Following \cite{Kalina2021}, we first require that galaxies have \texttt{FLAG\_GALFIT=2}, which means that the \textsc{GalfitM} fits have been successfully executed. If this flag is not equal to two, the fits were either not started because the specific object was not imaged in a sufficient number of bands or the fit did converge on a solution. For all fields, we require that all objects are imaged in at least three bands to ensure that objects with limited data are not fit. We note here that this requirement does not imply that objects must be detected in a minimum of three bands. In addition, we require that each source is imaged in the two filters closest to the rest-frame wavelengths at which we measure sizes (i.e.~1500{\AA} and 5000{\AA}) to avoid extrapolating fit parameters to wavelengths that are not sampled.

We also apply limits on redshift. Our redshift bins are selected to avoid using F275W photometry where the flux may be affected by Lyman break or Ly$\alpha$ effects. Specifically, we exclude the redshift range $1.3 < z < 1.5$, where we expect to observe almost no flux in F275W despite the filter's wide wavelength coverage. We additionally limit our sample to $z\geq0.5$ as the F275W filter does not sample 1500{\AA} below this redshift. Finally we exclude galaxies at $z>3.0$ as the rest-frame optical begins to fall redward of the H-band, the reddest filter used in this study. We note that other HST studies that have measured galaxy sizes in the rest-frame optical have also commonly extended their analyses out to $z\sim3$ \citep[e.g.][]{Buitrago2008ApJ, vdWel2014, Mowla2019ApJ}.

 
We further remove any objects that are likely stars by requiring that the \texttt{CLASS\_STAR} flag derived from \textsc{SExtractor} be less than 0.9, above which objects are likely point sources. We also require that all galaxies in our sample are one magnitude brighter than the 5$\sigma$ depth to ensure that all galaxies are bright enough to be modeled accurately. We discuss this selection in more detail in \S\ref{sec:results} as well as any selection biases that may impact our results. We further exclude any objects with magnitude differences greater than 0.5 between the UVCANDELS photometric catalogs and our \textsc{GalfitM} fits, as previously discussed. Finally, following \cite{Kalina2021}, we require that the effective radius satisfies 0.305 pixels $\leq$ Re $\leq$ 395 pixels in all bands to remove any objects that are too small to be modeled or are fit with unphysically large sizes. Similarly, the Sérsic index needs to satisfy 0.205 $\leq$ $n$ $\leq$ 8 in all bands, to exclude fits which have run into fitting constraints or have highly concentrated light profiles.

Lastly, for each object, the average Sérsic index from all bands is required to be $n\leq$2.5 such that we select disk-like systems as in e.g.~\cite{Trujillo2007MNRAS}. We note that this selection could be contaminated by galaxies that are not disk-like, especially at the highest redshifts where most galaxies have low Sérsic indices, including mergers and galaxies that are progenitors of ellipticals (see \citealt{Buitrago2013MNRAS} for further discussion). We find that using an average Sérsic index threshold of $n\leq$2.0 following e.g.~\cite{Ravindranath2004ApJ} changes the slopes of the best-fitting stellar mass--size relations by $\sim0.1\sigma$, at most. Thus the resulting best-fitting stellar mass--size relations are in very good agreement with those obtained when using the $n\leq$2.5 threshold. Moreover, the lower Sérsic index cut does not robustly allow us to remove galaxies that fall in the locus of the size-mass plane where quiescent galaxies are typically found. Combined, these results suggest that this selection does not impact the main results of this work and we therefore choose the more lenient cut of $n\leq$2.5.

\subsection{Simulated Data}\label{sec:simulateddata}
\begin{figure*}
    \centering
    \includegraphics[width=.98\textwidth]{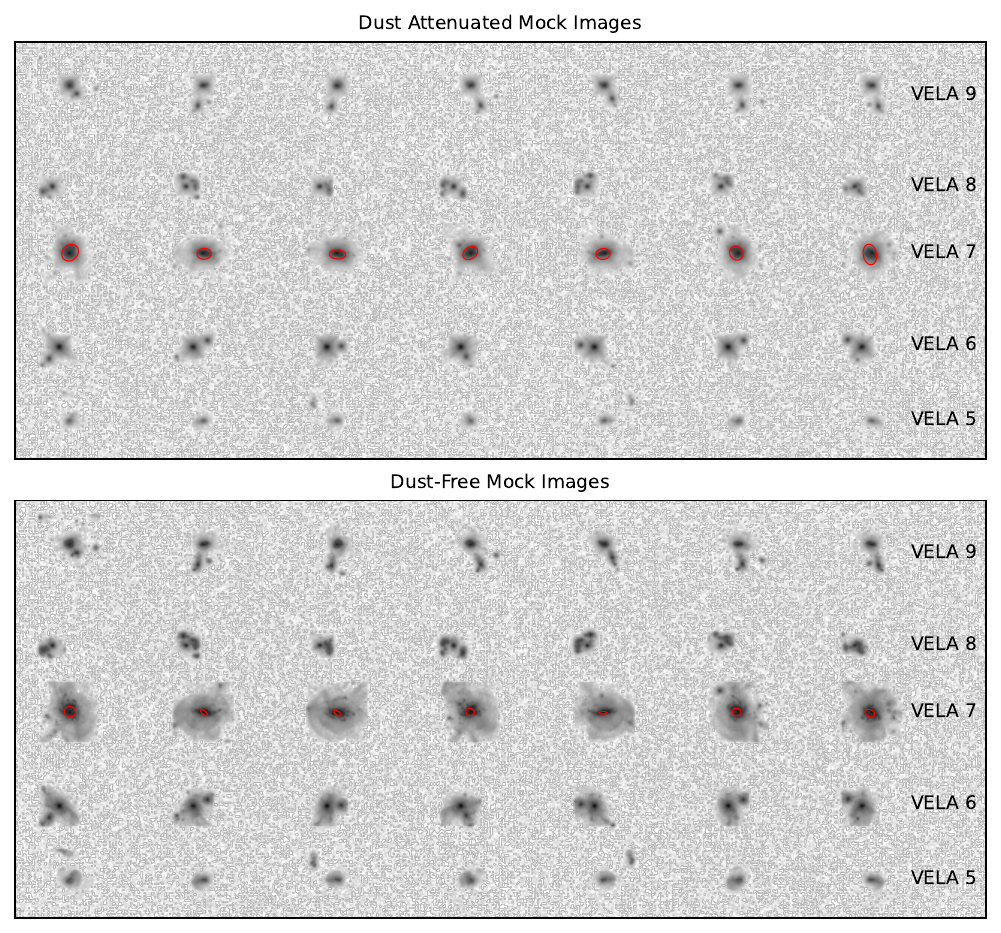}
    \caption{Example mock galaxy F160W images that we use as part of the input to \textsc{GalfitM}. The top panel shows galaxies where the full dust model is employed, while the bottom panel shows galaxies as they would appear without any diffuse dust. In both panels, each row shows individual VELA galaxies, where their ID numbers are labeled, viewed from seven different orientations -- face on, edge on, at a 45$^\circ$ inclination, and four random orientations -- shown in each column, from left to right. Visually, the dust-free galaxies generally appear brighter and more extended, although their half-light radii are smaller. Ellipses that contain half of the total flux, as derived from \textsc{GalfitM}, are shown in red for the most extended galaxy, VELA 7.} 
    \label{fig:example_mock}
\end{figure*}

As highlighted in \S \ref{sec:intro}, it is crucial to understand how the distribution of dust within galaxies affects the sizes that we measure at different wavelengths. 
To this end, we use a sample of synthetic images of 33 mock galaxies from the VELA simulation suite \citep{Ceverino2014MNRAS} that are available for a variety of viewing angles including face-on, 
edge-on, and random viewing orientations \citep{Simons2019ApJ}. 
We refer the reader to \cite{Ceverino2014MNRAS} for details about the simulation suite and to \cite{Simons2019ApJ} for the creation of the synthetic images, but we briefly address the elements that are relevant for this work below.


\cite{Simons2019ApJ} provide synthetic images of the generation 3 VELA galaxies on MAST\footnote{\url{https://archive.stsci.edu/prepds/vela} }, where the data products are created using the dust radiative transfer code, \textsc{Sunrise} \citep{Jonsson2006MNRAS.372....2J, Jonsson2010MNRAS,Jonsson2010NewA}. In this work, we use mock images for the generation 6, or VELA-6, simulations, which are created using an approach that is consistent with \cite{Simons2019ApJ}. 
These data are first described in \cite{Ceverino2023MNRAS} and have the same initial conditions as in the previous suite but a different feedback model. Mock images for these simulations are available in the observed frame in the F435W, F606W, F814W, F125W, F140W, and F160W filters, and include both images without the effects of dust (i.e.~images of the intrinsic dynamical properties) and with the SUNRISE dust-radiative transfer program. The differences between mock images where dust attenuation is and is not included can be seen in the left and right panels of Figure~\ref{fig:example_mock}, respectively. 

In order to provide a consistent comparison between the UVCANDELS data and the mock images, we match the properties of the simulated images to the data as much as possible. Specifically, we first downsample the simulated images to a pixel scale of 0\farcs03 per pixel and then convolve them with our PSFs, shown in the top row of Figure \ref{fig:PSFs+kernels}. 
We then add Poisson noise as well as noise that was quantified by sampling the noise distribution of our UVCANDELS data using a Monte Carlo technique. The latter accounts for read noise, dark current, and sky background.

Roughly 1/6 of the final synthetic galaxies can be seen in Figure~\ref{fig:example_mock}, where each row shows a different system of objects at $z=2$. We only study the properties of the simulated galaxies at this redshift because F435W is the bluest filter at which these mock images were created. 
From left to right, we include face-on, edge-on, 45$^\circ$ between face-on and edge-on, and four truly random orientations. The objects in the image on the top are dust attenuated while on the bottom, they are shown as they would appear without the presence of any dust. We note that visual impression would suggest that the galaxies without dust, shown on the bottom, are larger in size. However, the presence of dust flattens the light profile in the center such that the Sérsic index of the profile decreases. The \textit{a priori} expectation would therefore be that the dust-free galaxies would have smaller half-light radii (despite appearing more extended in Figure~\ref{fig:example_mock}) as their light profiles are more centrally concentrated. For VELA 7, which is the most extended mock galaxy shown in Figure~\ref{fig:example_mock}, we show ellipses that contain half of the total flux, as derived from \textsc{GalfitM}. In the top panel, where the effects of dust are included, it can be seen that the half-light radius is indeed larger for all orientations. 

Using these mock images, we measure the physical properties of the simulated galaxies using the Mega-Morph tools in a way that is consistent with the modeling performed on the UVCANDELS data. One difference is that we use the synthetic F160W image as our detection image, instead of constructing a white light image. 
This is done because we obtain stellar masses for the individual mock galaxies using \textsc{Dense Basis}, consistent with UVCANDELS. We do not use the true stellar masses from the simulations because these are provided for the whole system in each snapshot, which as shown in Figure~\ref{fig:example_mock}, can consist of multiple different objects. Hence, to be more consistent with the aperture photometry that was used to derive stellar masses for the UVCANDELS galaxies, we use the F160W image as our detection image. In practice, we find that the difference in galaxy sizes from using different detection images is small and not systematic and therefore, this choice does not strongly impact our results.

Finally, we apply the same selection criteria discussed in \S\ref{sec:uvcandels} to the mock objects, resulting in a sample of 205 dust-free and 168 dust attenuated mock galaxies. We highlight that we recover more dust-free mock galaxies as they are brighter.


\section{The evolution of the UV stellar mass--size relation of disk galaxies from UVCANDELS} \label{sec:results}
We have measured rest-frame UV and optical half-light radii for 15,996 disk galaxies from the GOODS-N, GOODS-S, EGS, and COSMOS fields. Throughout the remainder of the paper, we refer to the half-light radius along the semi-major axis as a galaxy's size, unless otherwise specified. We note here that these measurements are less affected by inclination effects than commonly used circularized half-light radii (see also \citealt{vdWel2014}). As \textsc{GalfitM} measures and fits galaxy sizes as a function of wavelength, we obtain rest-frame UV sizes at 1500{\AA} and rest-frame optical sizes at 5000{\AA} from the size polynomial directly.

\begin{figure*}
    \centering
    \includegraphics[width=1\textwidth]{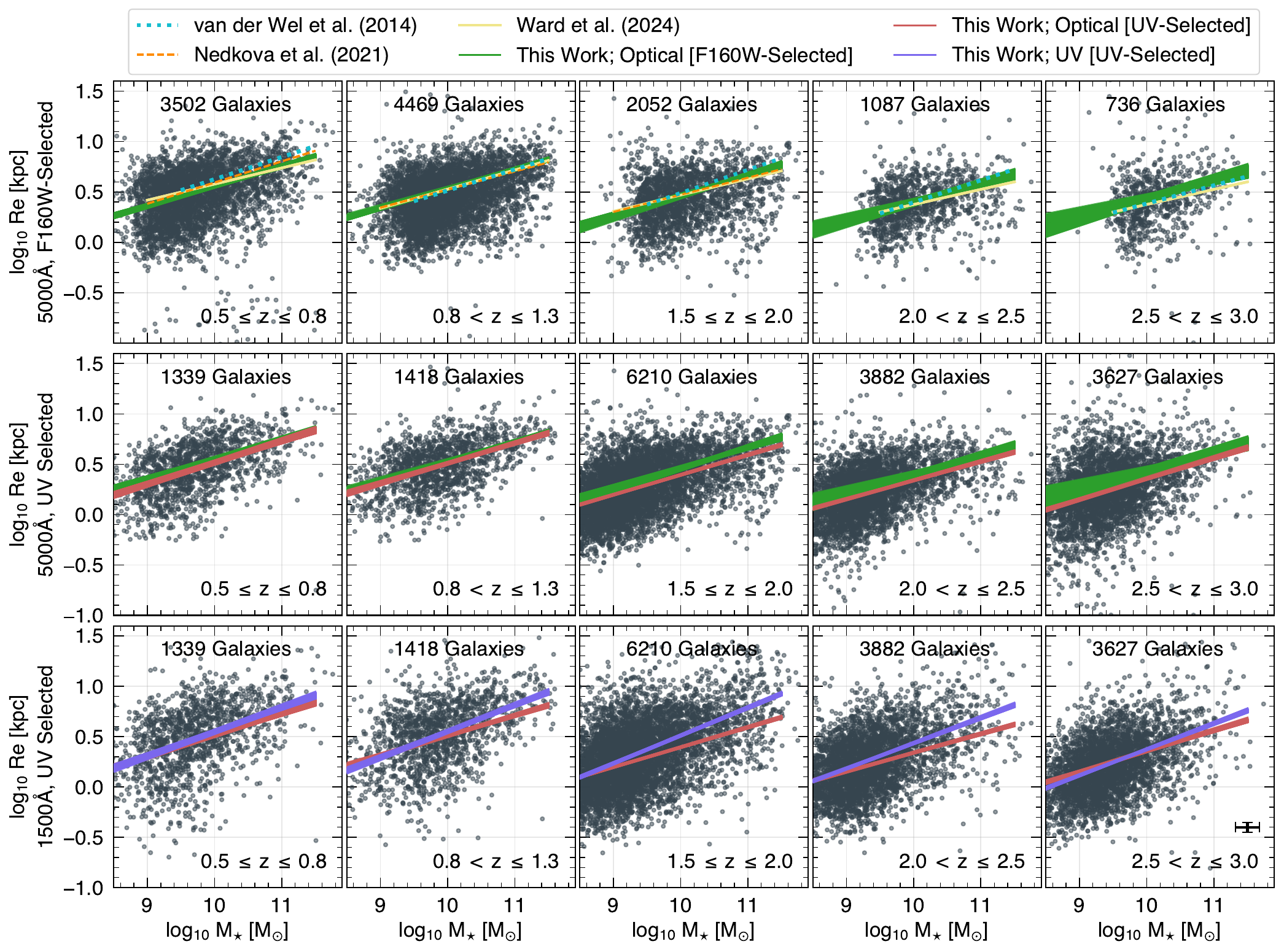}
    \caption{Stellar mass -- size relation of late-type galaxies, where the widths of the best-fitting relations, derived from Equation~\ref{eq:singlepl}, represent all fits with parameters that fall within 1$\sigma$ of the best-fit parameters. \textit{Top Row:} Optical stellar mass--size relations measured at 5000Å for galaxies that are selected to be one magnitude brighter than the 90\% completeness limit in F160W. These are shown to provide direct comparisons with the literature. \textit{Middle Row:} Same as the top row, but where galaxies are selected to be one magnitude brighter than the $5\sigma$ depth in F275W at $z\leq1.3$ and in F435W at $z\geq1.5$. The best-fitting relations for this sample are shown in red and the green relations are reproduced from the top row for comparison. \textit{Bottom Row:} The stellar mass -- size relations in the UV measured at 1500{\AA} for the same galaxies as in the middle row, that were selected via the same criteria. The bottom right panel shows a typical uncertainty for individual objects in our highest redshift bin.}
    \label{fig:mre}
\end{figure*}

\begingroup
\setlength{\tabcolsep}{5pt} 
\renewcommand{\arraystretch}{1} 

\begin{table*}
\centering
\caption{Best-fitting parameters for disk-like galaxies from the single power law shown in Equation~\ref{eq:singlepl} for the F160W selected optical relation and the UV-selected optical and UV stellar mass-size relations. For each, $\log_{10}$(A) is the intercept of the relation at M$_\star=5\times10^{10}$M$_\odot$ and B is the slope. The `F160W Selected Optical Relation' parameters correspond to the green lines in Figure \ref{fig:mre}, the `UV Selected Optical Relation' and `UV Selected UV Relation' parameters correspond to the red and blue lines, respectively. We report the 1$\sigma$ range as the uncertainty. The corner plots for all redshift bins are shown in Figure \ref{fig:cornerplots}. The rightmost column shows the significance of the difference between the UV Selected rest-frame optical and UV relations. }
\vspace{-0.4cm}
\begin{tabular}{  c | c c|c c |c c | c}
\hline
\multicolumn{1}{c |}{ \ } & 
\multicolumn{2}{c |}{F160W Selected Optical Relation} &
\multicolumn{2}{c |}{UV Selected Optical Relation} & 
\multicolumn{2}{c |}{UV Selected UV Relation} &
\multicolumn{1}{c }{ $\Delta$ Slope }\\ 

$z$ & $\log_{10}$(A) & B &  $\log_{10}$(A) & B& $\log_{10}$(A) & B  & Significance\\
\hline 

0.5$\leq$ $z$ $\leq$0.8 &
$0.692{\pm0.013}$&$0.196{\pm0.010}$&
$0.661$$\pm$$0.016$ & $0.212$$\pm$$0.013$ & $0.713$$\pm$$0.016$ & $0.237$$\pm$$0.013$ & 1.4$\sigma$\\ 

0.8$<$ $z$ $\leq$1.3 & 
$0.672{\pm0.011}$ & $0.193{\pm0.009}$ & 
$0.647$$\pm$$0.014$ & $0.200$$\pm$$0.013$ &
$0.721$$\pm$$0.014$ & $0.257$$\pm$$0.015$ & 3.5$\sigma$\\ 

1.5$\leq z\leq$2.0 &
$0.606{\pm0.013}$ & $0.207{\pm0.012}$ & 
$0.531$$\pm$$0.009$ & $0.197$$\pm$$0.006$ &
$0.694$$\pm$$0.009$ & $0.275$$\pm$$0.006$ & 9.2$\sigma$\\ 

2.0$<z\leq$2.5 & 
$0.525{\pm0.015}$ & $0.181{\pm0.015}$ & 
$0.469$$\pm$$0.012$ & $0.186$$\pm$$0.008$&
$0.608$$\pm$$0.013$ & $0.250$$\pm$$0.009$ & 5.3$\sigma$\\  

2.5$<z\leq$3.0 & 
$0.559{\pm0.022}$ & $0.178{\pm0.032}$ & 
$0.495$$\pm$$0.013$ & $0.203$$\pm$$0.010$ &
$0.550$$\pm$$0.014$ & $0.256$$\pm$$0.010$ & 4.1$\sigma$ \\  

\hline
\end{tabular}
\label{tab:from_eq2}
\end{table*}
\endgroup


\subsection{UV and Optical Stellar Mass-Size Relations} \label{sec:MRe}

In order to provide direct comparisons with previous results, we first select galaxies that are one magnitude brighter than the 90\% completeness limit in the F160W filter. Specifically, we use the limits reported by \cite{Skelton2014ApJS} and remove objects fainter than 24.1mag in the H-band. The stellar mass-size relations for this sample are shown in the top row of Figure~\ref{fig:mre}, where the dotted cyan, dashed orange, and solid yellow lines indicate the best-fitting results for star-forming galaxies from \cite{vdWel2014}, \cite{Kalina2021}, and \cite{Ward2024ApJ}, respectively.

For all stellar mass--size relations shown in Figure~\ref{fig:mre}, the best-fitting models with the 1$\sigma$ confidence ranges are derived by fitting the data with
 
 \begin{equation}
    R_e = A \Big( \frac{M_*}{5 \times 10^{10} \ \mathrm{M}_\odot} \Big) ^ {B},
    \label{eq:singlepl}
\end{equation}
\noindent which is commonly adopted in the literature for star-forming galaxies \citep[e.g.][]{ vdWel2014, Dimauro2019, Kalina2021, Ward2024ApJ, Ito2024ApJ}. Here, $R_e$ is the effective radius, M$_*$ is the stellar mass and $A$ and $B$ are fit parameters, describing the normalization at M$_\star=5\times10^{10}$M$_\odot$ and slope of the relation, respectively. These best-fitting parameters are obtained with a fitting approach that explores the parameter space with a Markov Chain Monte Carlo (MCMC) algorithm using the \textsc{emcee} package \citep{emcee2013PASP}. We assume uniform priors for all fit parameters when computing the likelihood function, which has the form:

\begin{equation}\label{eq:2}
    \ln \mathscr{L} = K + \frac{1}{\sqrt{2 \ \pi \ \Sigma_n^2}} \ \mathrm{exp}\Big(-\frac{\Delta_n^2}{2 \ \Sigma_n^2} \Big) \ \mathrm{where} \\
\end{equation}
 \begin{equation}\label{eq:3}
    \Delta_n = y_n - Bx_n - \log_{10}{A} + B\log_{10}(5\times10^{10}) \ \mathrm{and}
\end{equation}
\begin{equation}\label{eq:4}
    \Sigma_n^2 = \vec{v}^\mathrm{T}S_n \vec{v} \ \mathrm{where}
\end{equation}
\begin{equation}\label{eq:5}
    \vec{v}^\mathrm{T} = (-B, 1) \ \mathrm{and } \ S_n = w \begin{pmatrix}
\sigma_{x,n}^2 & \sigma_{xy,n}\\
\sigma_{xy,n} & \sigma_{y,n}^2
\end{pmatrix} 
\end{equation}

These equations are largely motivated by \cite{Hogg2010} and the \textsc{emcee} package documentation\footnote{Useful descriptions and code examples can be found at \url{https://emcee.readthedocs.io/en/stable/tutorials/line/} and \url{https://github.com/dfm-io/post--fitting-a-plane}}. 
We refer the reader to these works for the derivation of these equations and more details, but we briefly discuss some of the key aspects below.

We begin with the likelihood function shown in Equation \ref{eq:2}, where K is some constant, which can be ignored when maximizing the likelihood. The other parameters needed to calculate the likelihood are defined in Equations \ref{eq:3} and \ref{eq:4}. Equation \ref{eq:3} describes the difference between the n-th y-value and model evaluated at the n-th x-value. Here, $A$ and $B$ are the fit parameters from Equation \ref{eq:singlepl}. Finally, we describe the uncertainty tensor, $S_n$, defined in Equation \ref{eq:5} which allows us to take into account uncertainties in both mass and size when computing the likelihood function. 
In short, the size uncertainties are obtained from \textsc{GalfitM} and increased by a factor of three following \cite{Kalina2021} as the uncertainties from \textsc{GalfitM} do not reflect the
full uncertainty in the final fits \citep[see e.g.][]{MegaMorph, vdWel2014, Lange2016}. The uncertainty on the corrected stellar mass is taken to be the one derived with \textsc{Dense Basis}. Finally, we scale the uncertainty tensor by $w$, which is a normalized number density such that more massive galaxies, despite being a minority, play a key role in determining the slope of the stellar mass--size relation. The number density is calculated in bins of 0.2dex in stellar mass. We then normalize the number density in each bin by the bin with the highest number density. This effectively weights galaxies' contribution to the best-fitting relation by the inverse of the number density as a function of stellar mass. We find that applying this scaling factor does not significantly impact the optical stellar mass--size relations but slightly steepens the rest-frame UV relations. We note that the UV stellar mass--size relations are steeper than the optical ones even when this scaling factor is not applied.

Our best-fitting relations for the F160W-selected sample are shown in green in the top panels of Figure~\ref{fig:mre}. The width of the relations encompasses all possible fits with parameters that fall within $1\sigma$ of the best-fitting parameters. In general, we find good agreement with previous studies, although we note that our redshift bins are different from those commonly used in the literature. As discussed in \S\ref{sec:subselection}, objects at $1.3 <  z < 1.5$ are excluded in an effort to avoid galaxies for which the F275W detections are affected by the Lyman break. As a result, our first two redshift bins (i.e.~$0.5\leq z \leq 0.8$ and $0.8< z \leq 1.3$) cover a smaller redshift range. Although we do not correct for this difference, we find that our relations agree well with those from previous studies.

Another significant difference is that most previous works select their samples based on galaxy rest-frame colors using UVJ diagrams. Our approach of selecting disk-like galaxies using a Sérsic index cut of $n\leq 2.5$ will naturally result in a different selection. However, \cite{Kalina2021} have shown that sample selections based on rest-frame galaxy colors, specific star formation rates, and Sérsic indices all reproduce the same general trends in the stellar mass--size relation. Therefore it is not surprising that even though our selection is inherently different from those used in \cite{vdWel2014}, \cite{Kalina2021}, and \cite{Ward2024ApJ}, we still recover stellar mass--size relations that are consistent with the literature.

The galaxies shown in the middle and bottom rows of Figure~\ref{fig:mre} are selected to be one magnitude brighter than the $5\sigma$ depth in the filter closest to the rest-frame 1500{\AA}. Specifically, at $0.5 \leq z \leq 1.3$, we
exclude objects fainter than 26mag in F275W. At higher redshifts, we exclude galaxies fainter than 27mag in the F435W filter in the EGS and COMSOS fields. B-band imaging was not obtained for the GOODS-N and GOODS-S fields as part of the UVCANDELS program. Therefore, we use the existing B-band imaging in these fields and we select galaxies brighter than 26.7mag in the F435W filter based on the $5\sigma$ depths reported in \cite{Grogin2011ApJS..197...35G}. We find that this selection roughly corresponds to applying an SNR cut of $\sim$13, which is sufficient for morphological analyses \citep[][]{Ravindranath2004ApJ}. Lastly, we also remove galaxies that have an SNR that is lower than 13 in the filter closest to the rest-frame 5000{\AA}. This last selection removes 159 galaxies and does not alter the findings of this paper but it ensures that fit parameters are not extrapolated for galaxies that are not detected with a sufficient SNR in red filters. The final number of galaxies is indicated at the top of each panel in Figure~\ref{fig:mre}.

The detection rate of low-mass galaxies depends on their stellar mass-to-light ratios and redshifts, with cosmological surface brightness dimming playing an increasingly significant role with increasing redshift. In this work, our analyses include galaxies with stellar masses as low as $10^{8.5}$M$_\odot$ out to $z=3$. The CANDELS dataset is mass complete to $\lesssim$$10^{10}$M$_\odot$ at our highest redshift for star-forming galaxies \citep{vdWel2014} but we include less massive galaxies as \cite{Kalina2021} have shown that reliable galaxy sizes can be measured for mass-incomplete samples. However, they note that magnitude-limited samples are particularly biased against faint extended sources \citep[see also][]{Kramer2022ApJ, Windhorst2023AJ}, such as low surface brightness or ultra diffuse galaxies. These galaxies tend to be spheroidal or elliptical in structure \citep{Conselice2018RNAAS} and should therefore be largely removed from our sample anyway as we focus on disk-like galaxies in this study. As a result, any biases against large faint sources should not impact our main results.


From Figure~\ref{fig:mre}, we first note that at $z\leq1.3$, the UV-selected sample is smaller than the F160W-selected sample. This effect is unsurprising as galaxies are typically red and hence will appear fainter in F275W than in F160W. Therefore, we select more galaxies using the F160W magnitude cut despite having deeper UV observations. Conversely, at $z\geq1.5$, the UV-selected sample is significantly larger than the F160W-selected sample. At these redshifts, the UV-selected sample is obtained using a magnitude cut in the F435W filter. Since high redshift galaxies are fainter, we are able to obtain a larger UV-selected sample using the deeper UVCANDELS imaging.

From the middle row of Figure~\ref{fig:mre}, we find that the rest-frame optical stellar mass-size relations of the F160W- and UV-selected samples are fairly consistent, as indicated by the agreement between the red and green best-fitting relations. However, from the bottom row, where we compare the UV-selected relations in the rest-frame optical (shown in red) and the rest-frame UV (shown in blue), we find that the stellar mass--size relation in the UV is steeper at all redshifts. The best-fitting parameters and their $1\sigma$ uncertainties are reported in Table~\ref{tab:from_eq2} for each panel. In the rightmost column, we report the significance of the difference between the slopes of the rest-frame UV and optical relations, showing that for all but our lowest redshift bin, this is a $>$ 3$\sigma$ result. 

\begin{figure}
    \centering
    \includegraphics[width=0.48\textwidth]{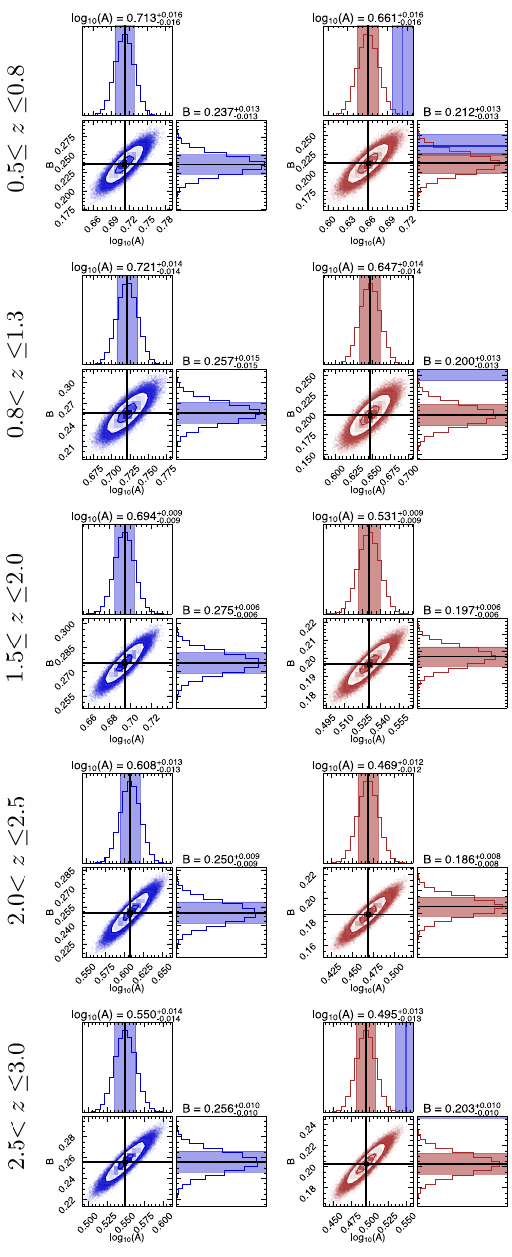}
    \vspace{-0.7cm}\caption{Corner plots of the best-fit parameters for the UV selected rest-frame UV (\textit{left}) and optical (\textit{right}) stellar mass--size relations. For the 1D posterior distributions, the shaded regions indicate $1\sigma$ confidence levels, shown in blue and red for the UV and optical parameters, respectively. The contours in the 2D posterior plots mark the 1$\sigma$ and 2$\sigma$ levels and the black lines indicate the best-fit parameters.}
    \label{fig:cornerplots}
\end{figure}

\begin{figure*}
    \centering
    \includegraphics[width=1\textwidth]{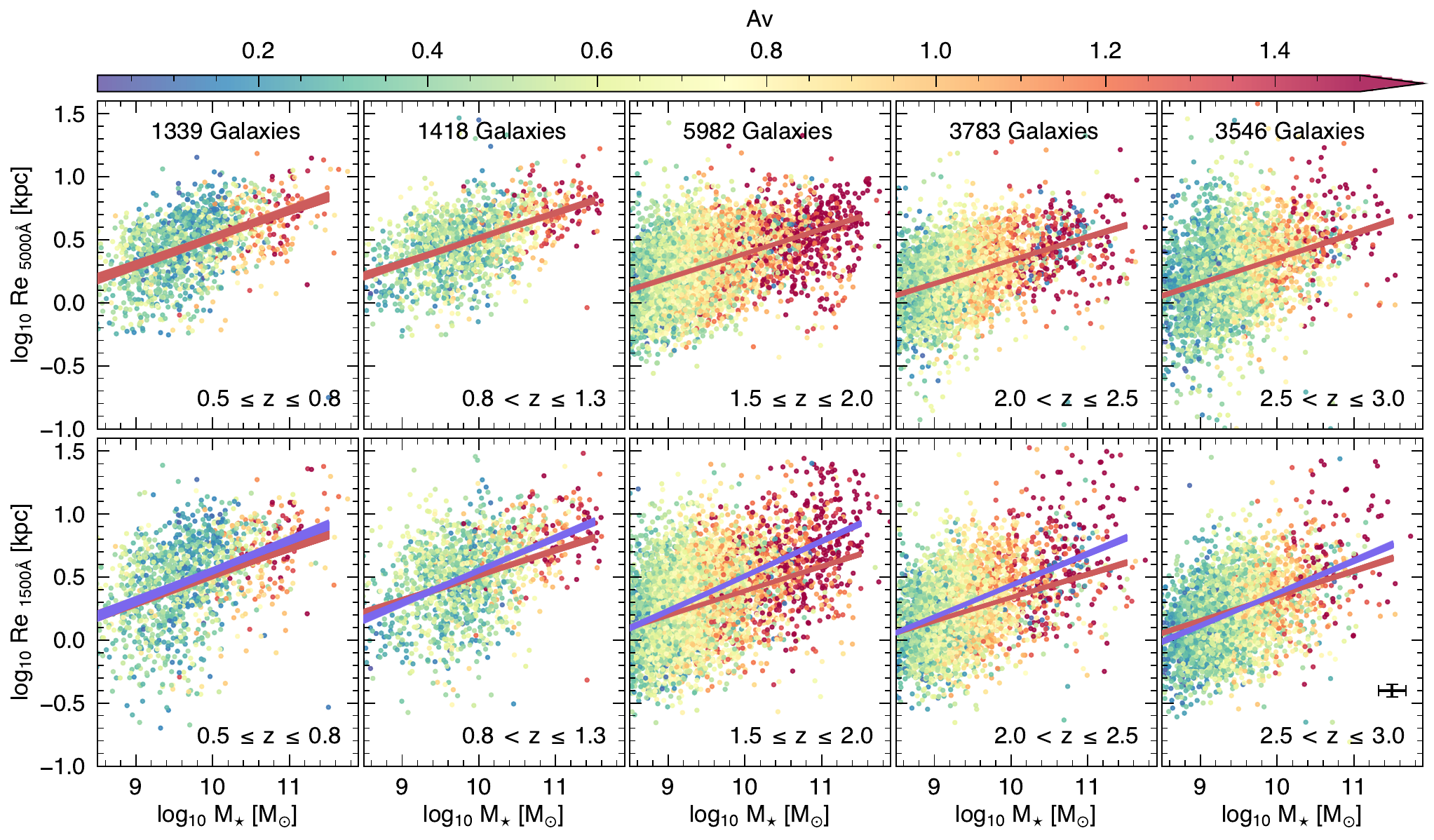}
    \caption{Stellar mass -- size relation of UV-selected disk galaxies, which are the same as in the middle and bottom rows of Figure~\ref{fig:mre}. Galaxies are color-coded by their dust attenuation (A$_\mathrm{V}$) as measured with \textsc{Dense Basis}. As expected, the most massive galaxies tend to be among the dustiest at all redshifts. The best-fit optical and UV relations are reproduced from Figure~\ref{fig:mre} in red and blue, respectively.}
    \label{fig:mre_av}
\end{figure*}

In Figure~\ref{fig:cornerplots}, we further compare the slopes by showing the resulting 1 and 2D posterior probabilities from our Bayesian fitting approach. The corner plots on the left show the results for the slope and intercept of the UV relations and the corner plots on the right show the best-fitting parameters for the optical rest-frame 5000{\AA} relations, with the UV 1D posterior probabilities overplotted for easier comparison. For the $1.5 \leq z \leq 2.0$ and $2.0 < z \leq 2.5$ redshift bins, both the slopes and intercepts of the rest-frame UV relations fall outside the range over which the probability distributions of the parameters for the rest-frame optical relation are shown.



The steeper relations in the UV suggest that the outskirts of disk-like galaxies in UVCANDELS are bluer in color than their inner regions. This result is not surprising for galaxies with central bulge and outer disk structures \citep[e.g.][]{Moorthy2006MNRAS, Morelli2008MNRAS.389..341M, Morelli2016MNRAS.463.4396M, Coccato2018MNRAS.477.1958C, Johnston2022MNRAS.514.6141J}. As previously discussed, one common interpretation has been that these color gradients indicate that disk galaxies are building up their stellar mass inside-out; however centrally concentrated dust can produce the same color gradients. Therefore, in what follows, we explore how dust attenuation is affecting our results.


\subsection{Impacts of Dust on UV and Optical Sizes} \label{sec:results_dust}

To evaluate whether dust may be responsible for the results presented in \S\ref{sec:MRe} to some extent, in Figure~\ref{fig:mre_av}, we show the same relations from the middle and bottom panels of Figure~\ref{fig:mre} but color-coded by dust attenuation (A$_\mathrm{V}$). We infer the dust attenuation with \textsc{Dense Basis}. The SED fitting process is briefly described in \S\ref{sec:photo_z_M*} and full details will be provided in Mehta et al.~(\textit{in prep}).

\begin{figure*}
    \centering
    \includegraphics[width=1\textwidth]{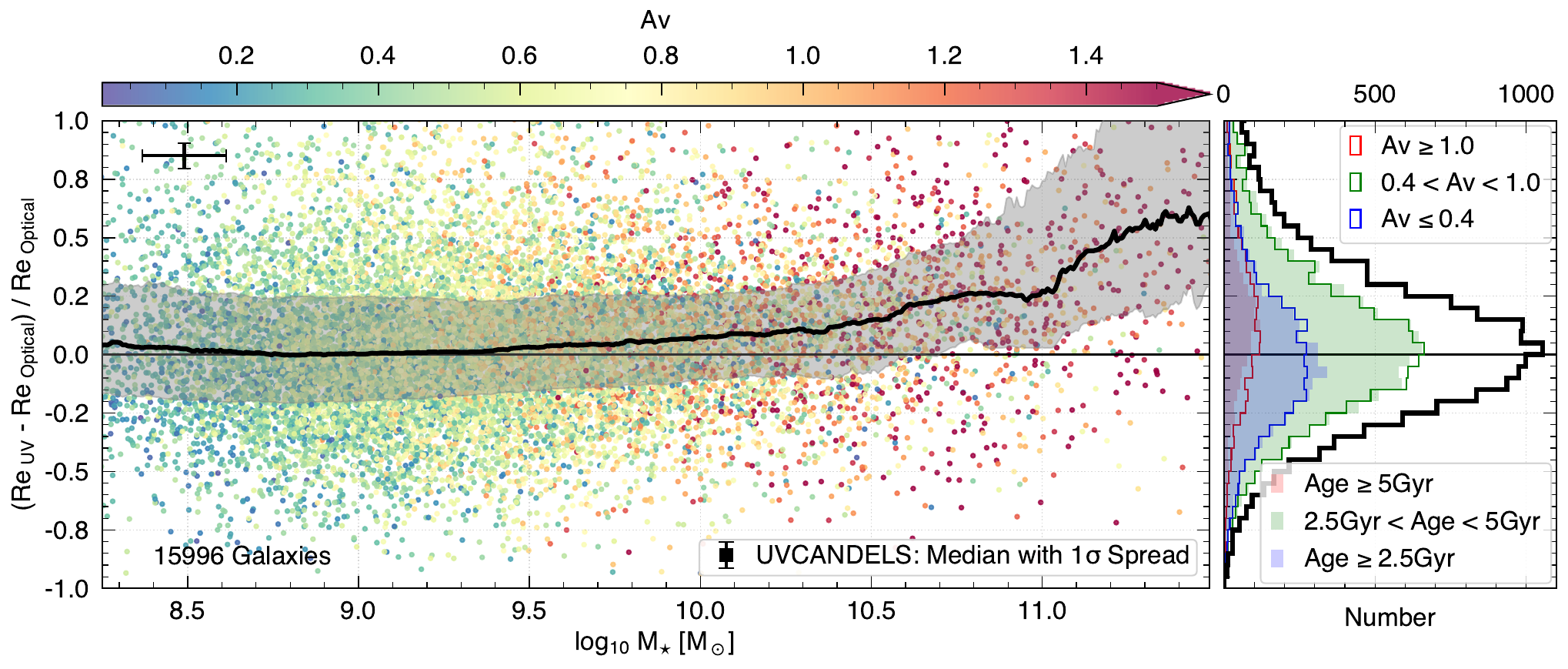}
    \caption{\textit{Left:} Difference between UV and optical sizes normalized by optical size as a function of stellar mass, where galaxies are color-coded by A$_\mathrm{V}$. Massive galaxies are often the most attenuated and are typically significantly larger in the UV than in the optical. This can be seen from the black line and gray shaded region, which represent the running median and 1$\sigma$ spread of the distribution. Typical size and mass uncertainties are shown in the upper left corner. \textit{Right:} Distribution of the rest-frame UV and optical size difference normalized by optical size for the total sample (thick black line), three bins of A$_\mathrm{V}$, and three bins of stellar age, as labeled.} 
    \label{fig:size_diff}
\end{figure*}


We first note that the most massive galaxies are more heavily attenuated than their lower mass counterparts, as expected from both observations \citep[e.g.][]{Martis2016, Whitaker2017ApJ...850..208W, Barger2023ApJ} and simulations \citep[e.g.][]{Cullen2017MNRAS, Wilkins2017MNRAS}. We also find that there are more galaxies with high levels of attenuation in the $1.5\leq z\leq2.0$ and $2.0< z\leq2.5$ redshift bins compared to our other redshift bins. This result is again unsurprising as we expect that the dust content in galaxies peaks at $z\sim2-3$ \citep{Madau}. It is interesting that these two redshift bins, where we observe the highest number of dusty galaxies, are also where we find the largest differences in the slopes of the best-fitting rest-frame UV and optical stellar mass--size relations, as indicated in Figure~\ref{fig:cornerplots} and Table~\ref{tab:from_eq2}. These results suggest that dust could be playing a significant role in the slope of the stellar mass--size relation of disk galaxies.


An important consideration here is that there is a well-known degeneracy between age and dust \citep[e.g.][]{Papovich2001ApJ}. 
This degeneracy can be broken by using infrared information \citep{Kauffmann2003MNRAS}. We hence emphasize that in our SED modeling of all four fields, we use ancillary \textit{Spitzer}/IRAC data (see \S\ref{sec:photo_z_M*} for more details). These are available in at least one channel for $\sim61$\% of our sample, thereby constraining the dust luminosity of most galaxies. The fraction of objects with \textit{Spitzer}/IRAC imaging increases to $\sim94$\% for galaxies with high stellar masses ($>10^{10}$M$_\odot$).





The difference between rest-frame UV and optical sizes as a function of stellar mass and dust attenuation for our UVCANDELS sample is presented in Figure~\ref{fig:size_diff}, where we combine our galaxy sample from all redshifts, and color-code each object by its A$_\mathrm{V}$ as in Figure~\ref{fig:mre_av}. The right panel of Figure~\ref{fig:size_diff} shows the size difference distribution normalized by the optical size for the total sample (thick black histogram), three bins of A$_\mathrm{V}$ (red, green, and blue histograms), and three bins of stellar age (filled red, green, and blue histograms). We obtain stellar ages, $t_{40}$, i.e.~the lookback time when 40\% of the stellar mass has been formed, from \textsc{Dense Basis}. 

From the right panel of Figure~\ref{fig:size_diff}, we first note that galaxies with high attenuation (A$_\mathrm{V}$ $\geq$ 1.0) and galaxies with older stellar populations ($t_{40}$ $\geq$ 5Gyr) typically have sizes that are larger in the rest-frame UV than in the optical. 
This is indicated by the two red histograms and suggests that the most massive galaxies are both more dust attenuated and host older stellar populations than lower mass galaxies. Second, this panel shows that the distribution of the UV and optical size difference normalized by the optical size peaks at zero for the full sample. This indicates that the majority of our galaxies are mid- to low-mass systems as the most massive galaxies tend to have larger rest-frame UV sizes as shown in the left panel, which we discuss in more detail below.

In the left panel of Figure~\ref{fig:size_diff}, the running median and $1\sigma$ spread are shown as a black line and gray shaded region, respectively, where each measurement is independent of the adjacent measurements such that our scatter is uncorrelated. Figure~\ref{fig:size_diff} shows that the rest-frame UV and optical sizes are consistent for lower mass galaxies ($\lesssim10^{9.5}$M$_\odot$), while higher mass galaxies are significantly larger in the rest-frame UV. Our results are in general agreement with \cite{Shibuya2015}, who also find that the difference between UV and optical sizes becomes significant for massive galaxies.

The interpretation of the results presented in Figure~\ref{fig:size_diff} is that only the more massive galaxies ($\gtrsim$$10^{10}$M$_\odot$), which are also more heavily attenuated and typically host older stellar populations, have bluer outskirts and redder centers. On average, the lower mass galaxies have similar sizes in the rest-frame UV and optical, suggesting that they do not have the same color gradients seen in the higher mass galaxies. As we do not observe strong color gradients in the low mass sample, which has lower levels of dust attenuation as shown in the left panel of Figure~\ref{fig:size_diff} and strong color gradients for the higher mass sample, we posit that the differences in sizes at different wavelengths are impacted by dust. Again, these results could suffer from the age-dust degeneracy; however, 94\% of our high-mass sample has available \textit{Spitzer}/IRAC imaging which helps break the degeneracy. Moreover, we find that when we limit our sample to disk-like galaxies that have both significant attenuation (A$_\mathrm{V}$ $>$ 0.6) and older stellar populations ($t_{40}$ $\geq$ 3 Gyr) 
than the typical galaxy in our sample, we observe a median trend that is highly consistent with the one shown in the left panel of Figure~\ref{fig:size_diff}. This implies that the age-dust degeneracy has a negligible effect on our results as this population does not suffer from the degeneracy because these objects already have an old stellar population. We investigate the effects of dust further by using simulations in the following Section. 

\subsection{Spatial Dust Distribution in Simulations}\label{sec:results_dust_distribution}

    
    
\begin{figure}
    \centering
    \includegraphics[width=0.48\textwidth]{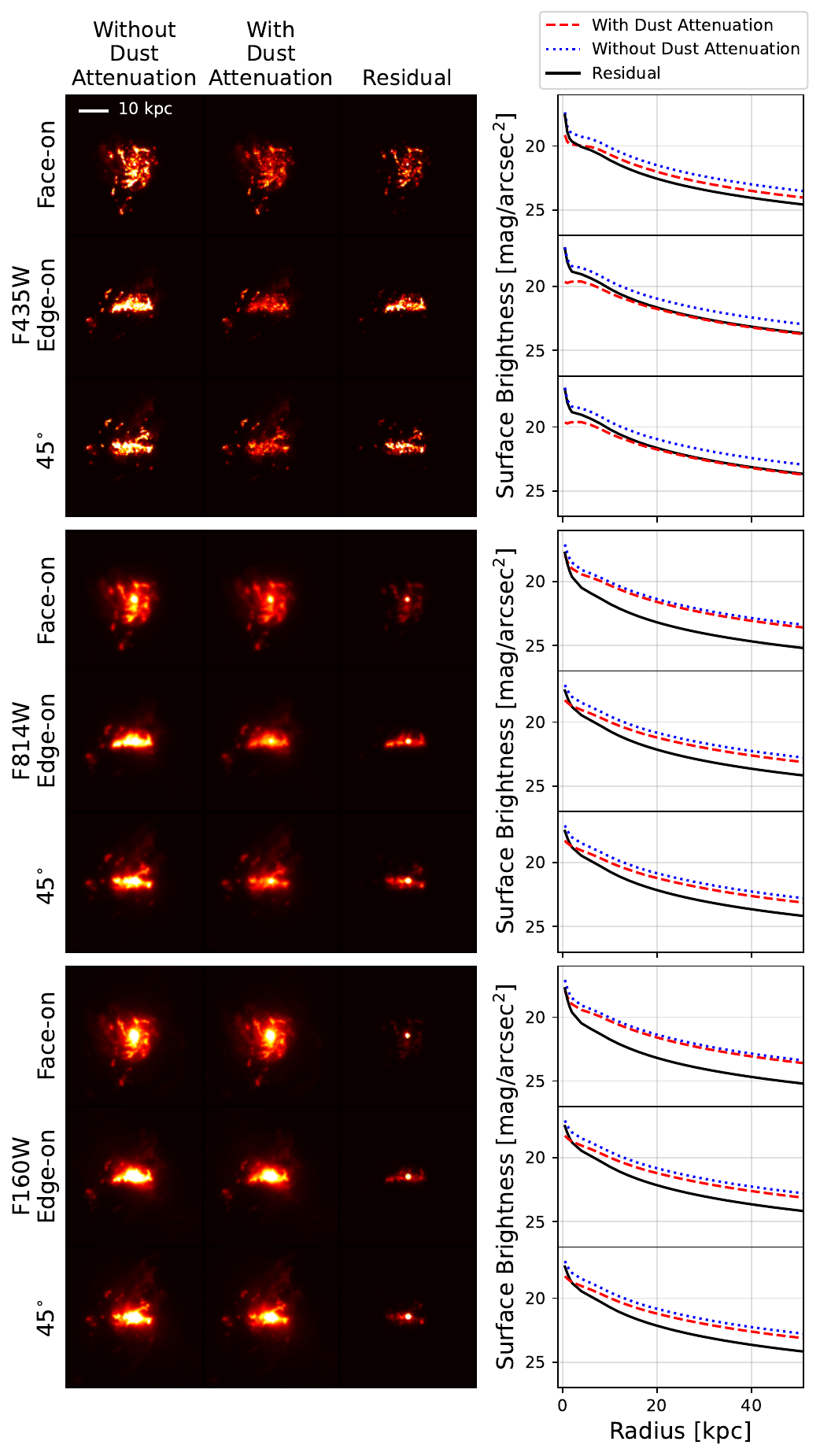}
    \caption{The VELA 1 galaxy, in  F435W (top nine images), F814W (middle nine images), and F160W (bottom nine images). For each filter, we show the object as it would appear without dust in the first column, with dust in the second column, and the difference (i.e.~residuals) in the third column for three different orientations. The residuals show the light that is attenuated such that this traces the distribution of dust in this galaxy. Finally, surface brightness profiles are shown in the rightmost column.}
    \label{fig:vela_dust_profiles}
\end{figure}

If dust is driving the difference in the stellar mass--size relations in the rest-frame UV and optical, then galaxies must be more attenuated in their central regions than in their outskirts. We test this by measuring the surface brightness profiles for our simulated galaxy sample. These are shown for one example galaxy at $z=2$ across three filters and three viewing orientations in Figure~\ref{fig:vela_dust_profiles}. In this figure, the first column shows the VELA 1 mock galaxy as it would appear without any dust attenuation and the second column shows the same mock galaxy when the full dust models are included. In the third column, we show the difference between these (i.e.~the residual), which can be interpreted as the amount of light that is attenuated by dust. Hence, the residual shows the spatial distribution of dust within this galaxy. All of these images have the same scale and are shown over the same dynamic range. Finally, the surface brightness profiles of each are shown in the rightmost column of Figure~\ref{fig:vela_dust_profiles}. At all orientations, the surface brightness profile of the residual (i.e.~the amount of light attenuated by dust), shown as a solid line in the rightmost panels is consistently brightest in the center. This shows that the central parts of the galaxy are the most heavily dust attenuated, and indeed we find this to be the case for all of the simulated galaxies. 

Centrally concentrated dust could be driving the differences between the rest-frame UV and optical stellar mass--size relations, although we note that the amount of light that is attenuated is dependent on the viewing angle. This can be seen for the example galaxy shown in Figure~\ref{fig:vela_dust_profiles} as well, where the surface brightness profile of the residual is faintest when the galaxy is viewed face-on. This suggests that less light is attenuated when galaxies are viewed face-on than at other orientations, as can be seen by comparing the black curve in the face-on case to the edge-on and 45$^\circ$ cases in all filters. We do not find a significant difference between the edge-on and 45$^\circ$ orientations.

Lastly, Figure \ref{fig:vela_dust_profiles} also shows that the amount of light that is attenuated by dust depends on the band at which the observation is made, as expected. In particular, the residuals, shown in the third column of Figure~\ref{fig:vela_dust_profiles} in the F435W band are brighter than in the other bands. This suggests that the B-band is more affected by dust than redder bands, with the H-band being the least affected. This is again expected as dust has a stronger effect at bluer wavelengths.





\section{Stellar Mass--Size Relations from Simulations} \label{sec:mre_from_sim}

\begin{figure*}
    \centering
    \includegraphics[width=\textwidth]{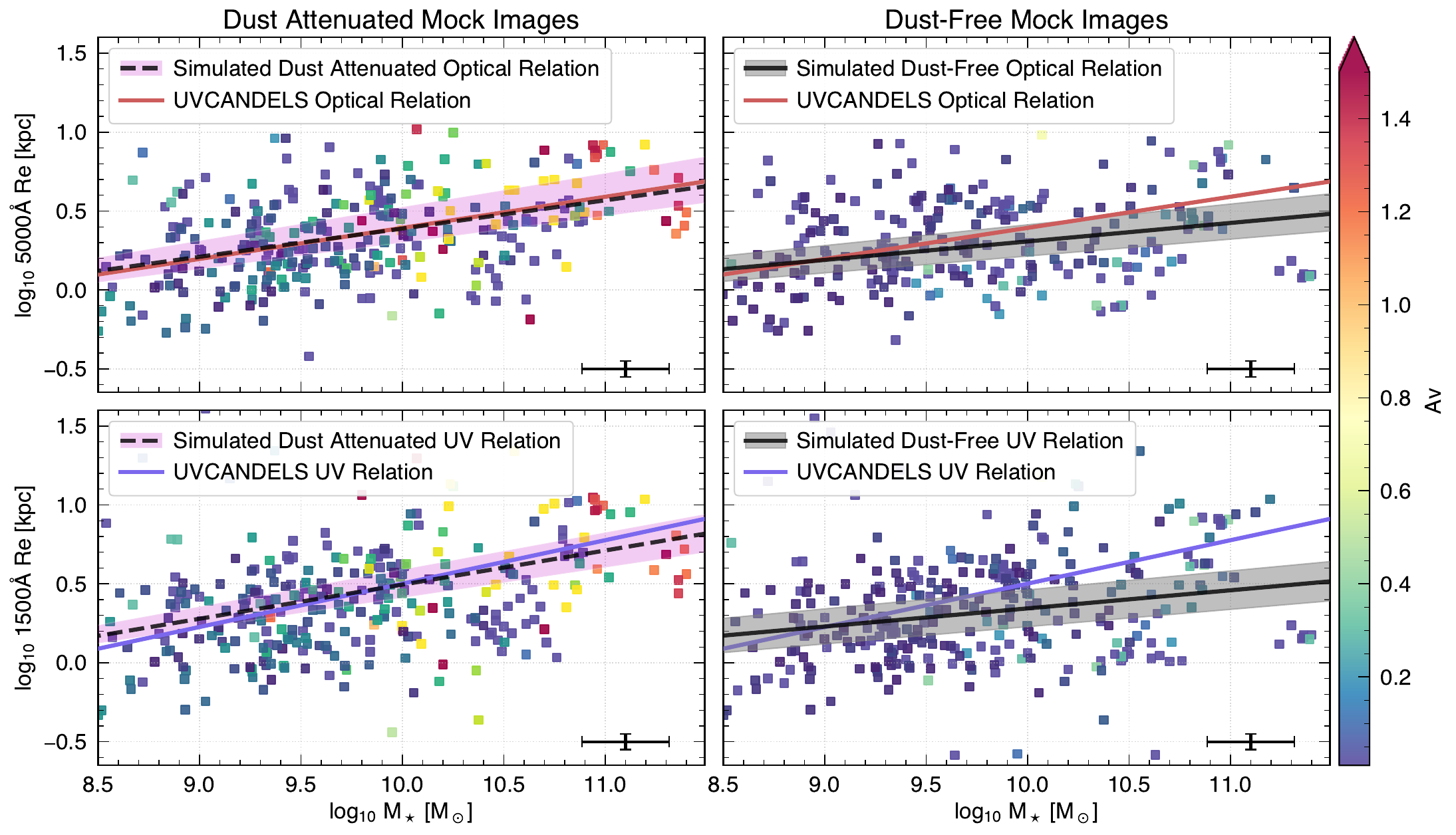}
    \caption{Stellar mass--size relations of the mock VELA galaxies at $z=2$. The panels on the left show the rest-frame optical (top) and UV (bottom) relations when the full dust prescription is included in the simulation and the panels on the right show the same but when no dust is present in the mock images. All simulated galaxies are color-coded by their dust attenuation, where galaxies in the left panels are on average more attenuated than galaxies in the right panels, as expected. The best-fitting stellar mass--size relations are shown in black, with the observed UV and optical relations over-plotted for comparison. }
    \label{fig:vela_MRe}
\end{figure*}

In \S \ref{sec:MRe}, we present the stellar mass--size relation of disk galaxies in the UVCANDELS fields in the rest-frame UV and optical, finding that the relations are significantly steeper in the UV. This result suggests that these galaxies have redder centers compared to their outskirts, consistent with numerous previous works \citep[e.g.][]{Shibuya2015, Suess2022ApJ}. We also find that the number of heavily attenuated galaxies peaks at $1.5 \lesssim z \lesssim 2.5$, which is also the redshift range at which we find the largest differences between the slopes of the rest-frame UV and optical stellar mass--size relations. In \S\ref{sec:results_dust}, we show that these trends are mainly driven by the most massive galaxies in our sample, which are also among the dustiest. Combined, these results suggests that dust attenuation is playing a significant role in the wavelength-dependence of the stellar mass--size relations that we observe.

In an effort to quantify the level to which dust attenuation is affecting the differences in size observed at different wavelengths, we turn to VELA, a set of cosmological zoom-in simulations with a comoving cosmological box of 40 Mpc/h \citep{Ceverino2014MNRAS}. One of the main features of these simulations is that mock images were created by including and excluding dust as discussed in \S\ref{sec:simulateddata}.  We refer the reader to 
\citealt{Simons2019ApJ} for full details, but in short, each simulation snapshot is post-processed with the dust radiative transfer software \textsc{Sunrise} \citep{Jonsson2006MNRAS.372....2J, Jonsson2010MNRAS,Jonsson2010NewA}.

The VELA mock images are also available at different viewing angles, as this likely affects the amount light that is attenuated by dust.
Using seven of these orientations for each system, we derive stellar mass--size relations for the dust-free and dust attenuated mock galaxies in a consistent way as for our UVCANDELS data. The results are shown in Figure~\ref{fig:vela_MRe}.

We show the stellar mass--size relation of mock disk galaxies at $z=2$ when the effects of dust are included in the left panels of Figure~\ref{fig:vela_MRe} and when they are not included in the right panels. All galaxies are color-coded by their dust attenuation as measured with \textsc{Dense Basis}. We measure the physical properties of our mock galaxy sample using the photometry as measured with \textsc{GalfitM} as well as the same software and assumptions as for the UVCANDELS data. As briefly discussed in \S\ref{sec:simulateddata}, we note that we cannot use the true mass and dust properties from the simulations as these are provided for the full system. At $z=2$, the snapshots of some systems are composed of multiple individual galaxies that by $z=0$ have merged to form a single star-forming galaxy. We therefore measure the physical properties of each mock galaxy individually in a way that is consistent with the UVCANDELS data. Additionally, we measure sizes for these objects using the same methodology for UVCANDELS, and apply the same selection criteria discussed in \S\ref{sec:selection}. Our final mock sample consists of 205 dust-free galaxies and 168 dust attenuated mock galaxies.

\begingroup
\setlength{\tabcolsep}{6pt} 
\renewcommand{\arraystretch}{1} 

\begin{table*}
\centering
\caption{The best-fitting parameters for the mock VELA galaxies using the single power law shown in Equation \ref{eq:singlepl}. As in Table~\ref{tab:from_eq2}, $\log_{10}$(A) is the intercept at M$_\star=5\times10^{10}$M$_\odot$ and B is the slope of the stellar mass--size relation of the mock galaxy sample. The uncertainties on each best-fit parameter represent the 1$\sigma$ confidence ranges. In the rightmost column, we report the significance of the slope difference between the stellar mass--size relation of the dust attenuated and dust-free mock samples, where the slope is shallower when dust is not included in the mock images. }
\begin{tabular}{c c|c c | c}
\hline
\multicolumn{2}{c |}{Optical Dust Attenuated} & 
\multicolumn{2}{c |}{Optical Dust-Free} &
\multicolumn{1}{c }{$\Delta$ Slope}\\
$\log_{10}$(A) & B &  $\log_{10}$(A) & B & Significance \\

$0.514$$\pm$$0.031$ & $0.179$$\pm$$0.028$ &
$0.389$$\pm$$0.021$ & $0.118$$\pm$$0.031$ & 1.5$\sigma$ \\ 

\hline
\hline
\multicolumn{2}{c |}{UV Dust Attenuated} &
\multicolumn{2}{c |}{UV Dust-Free } &
\multicolumn{1}{c }{$\Delta$ Slope}\\

$\log_{10}$(A) & B & $\log_{10}$(A) & B & Significance\\
\hline
$0.647$$\pm$$0.039$ & $0.216$$\pm$$0.038$ & 
$0.424{\pm0.029}$&$0.115{\pm0.029}$ &2.1$\sigma$ \\ 
\hline
\end{tabular}
\label{tab:mock_MRE}
\end{table*}
\endgroup

The positions of these mock galaxies on the size--mass plane are shown as squares in Figure~\ref{fig:vela_MRe}. In the top panels, we show the stellar mass--size relation in the rest-frame optical and in the bottom panels, we show the same but in the rest-frame UV. The best-fitting parameters for each relation that we derive for the mock galaxies using Equation~\ref{eq:singlepl} are given in Table~\ref{tab:mock_MRE}. In the figure, we have overlaid the best-fitting UV and optical stellar mass--size relation for UVCANDELS in blue and red, respectively, for easier comparison.

First, we note that galaxies in the left panels are more dust attenuated than galaxies from the dust-free mock images, as indicated by the color-coding. This is expected due to the nature of the two separate categories but it serves as a simple test of the SED modeling and physical parameter estimation. There are a few galaxies from the dust-free sample that are modeled with some attenuation; however, this is not surprising as 
there is some uncertainty associated with these measurements. 
We find that the mock sample has an average A$_\mathrm{V}$ uncertainty of $\sim$$0.14$ for the dust-free sample and $\sim$$0.39$ for the dust-attenuated galaxies. For reference the average A$_\mathrm{V}$ uncertainty on the UVCANDELS data is $\sim$$0.32$. The larger A$_\mathrm{V}$ uncertainty for the dust attenuated mock galaxies is likely due to fitting the SEDs with fewer bands as the UVCANDELS data have coverage in more filters. The dust-free mock sample has the smallest average A$_\mathrm{V}$ uncertainty because they are fit with the least amount of dust attenuation.

Second, we compare the simulated stellar mass--size relations to the observed UVCANDELS relations in each panel. We begin with the rest-frame optical shown in the top panels of Figure~\ref{fig:vela_MRe}. For all simulated relations, the best-fitting relations are shown in black and the magenta and gray shaded regions indicate all possible relations that have parameters, log$_{10}$(A) and B, that fall within $1\sigma$ of the best-fit parameters. 
In the top left panel, we show a comparison of the simulated and observed rest-frame optical stellar mass--size relations using the dust-attenuated mock images. The UVCANDELS optical relation is in excellent agreement with the relation of the simulated galaxies when full dust physics are included in the VELA simulation, and indeed the UVCANDELS relation falls within 1$\sigma$ across our entire mass range. For the dust-free mock galaxies in the top right panel of Figure~\ref{fig:vela_MRe}, we again compare the mock and UVCANDELS relations. Although the best-fitting relation is mostly consistent with the UVCANDELS optical relation at 10$^{10}$M$_\odot$, the rest-frame optical stellar mass--size relation for the dust-free mock galaxies is shallower.


Next, we compare the stellar mass--size relations in the rest-frame UV shown in the bottom panels of Figure~\ref{fig:vela_MRe}. As for the optical, we find that the stellar mass--size relation for the mock images where dust attenuation is included is in fair agreement with the rest-frame UV relation that we drive for UVCANDELS. This is shown in the bottom left panel of the figure and suggests that the VELA simulations are accurately reproducing the observed rest-frame optical and UV stellar mass--size relations. For the dust-free mock images shown in the right panels, we find that the simulated relation in the rest-frame UV is significantly shallower than the UV relation derived for UVCANDELS. We highlight that the difference in slope is smaller in the optical than it is in the UV, which is also indicated in the rightmost column of Table~\ref{tab:mock_MRE}.
This is consistent with the expectation that dust would impact rest-frame UV light more than in the rest-frame optical.

\begin{figure*}
    \centering
    \includegraphics[width=\textwidth]{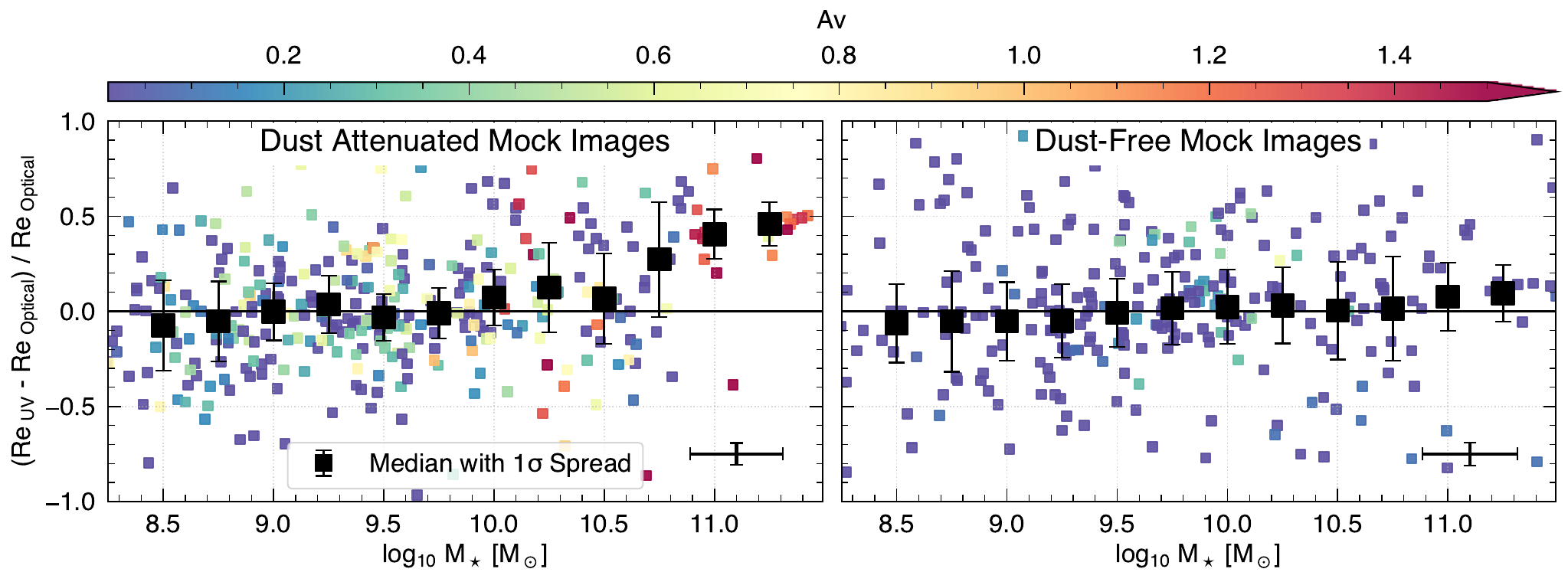}
    \vspace{-0.7cm}\caption{A similar comparison of rest-frame UV and optical sizes as in Figure~\ref{fig:size_diff} but for simulated VELA galaxies at $z=2$. In the left panel, we show size comparisons of simulated disk galaxies when dust is included. In general, high stellar mass galaxies are larger in the UV. The right panel shows the same but when the effects of dust are not included in the simulation. At all stellar masses, the difference between the rest-frame optical and UV sizes is consistent with zero when dust is not included. This demonstrates that dust alone can explain differences between the UV and optical sizes of massive galaxies.
    }
    \label{fig:simulation_size_diff}
\end{figure*}

Lastly, in Figure~\ref{fig:simulation_size_diff}, we compare UV and optical sizes of our mock galaxies in the same way as for the UVCANDELS sample in Figure~\ref{fig:size_diff}. Specifically, in the left panel, we show the difference between the rest-frame UV and optical sizes normalized by the optical size as a function of mass for the mock galaxy sample where the effects of dust are included. In the right panel, we show the same but for the dust-free case. All galaxies are color-coded by their dust attenuation, consistent with Figure~\ref{fig:vela_MRe}. The running median and $1\sigma$ spread of the distribution are shown as black squares and errorbars at 0.5dex intervals.

From the left panel of Figure~\ref{fig:simulation_size_diff}, we note that galaxies at $\sim10^{10}$M$_\odot$ begin to exhibit larger differences between their rest-frame UV and optical sizes, roughly consistent with what we find for UVCANDELS in Figure~\ref{fig:size_diff}. The right panel of Figure~\ref{fig:simulation_size_diff} shows that when the effects of dust are removed, disk galaxies have consistent rest-frame UV and optical sizes at all stellar masses. This would suggest that dust is causing more massive galaxies to be larger in the rest-frame UV by attenuating significant amounts of UV light from their central regions. These results and their implications are further discussed in the following section.

\section{Discussion} \label{sec:discussion}
We have presented stellar mass--size relations in the rest-frame UV and optical for UVCANDELS galaxies in \S\ref{sec:results} and mock galaxies from the VELA simulations in \S\ref{sec:mre_from_sim}. By comparing simulations to observations, we find that although the VELA simulations produce stellar mass--size relations that are in agreement with our UVCANDELS measurements when dust is included, the dust-free relations are significantly flatter. This directly shows that the presence of dust is steepening the slope of the stellar mass--size relation of disk galaxies. 
This steepening is more significant in the rest-frame UV than in the optical, as expected since the dust is centrally concentrated and it attenuates rest-frame UV light more than optical.

\begin{figure}
    \centering
    \includegraphics[width=0.47\textwidth]{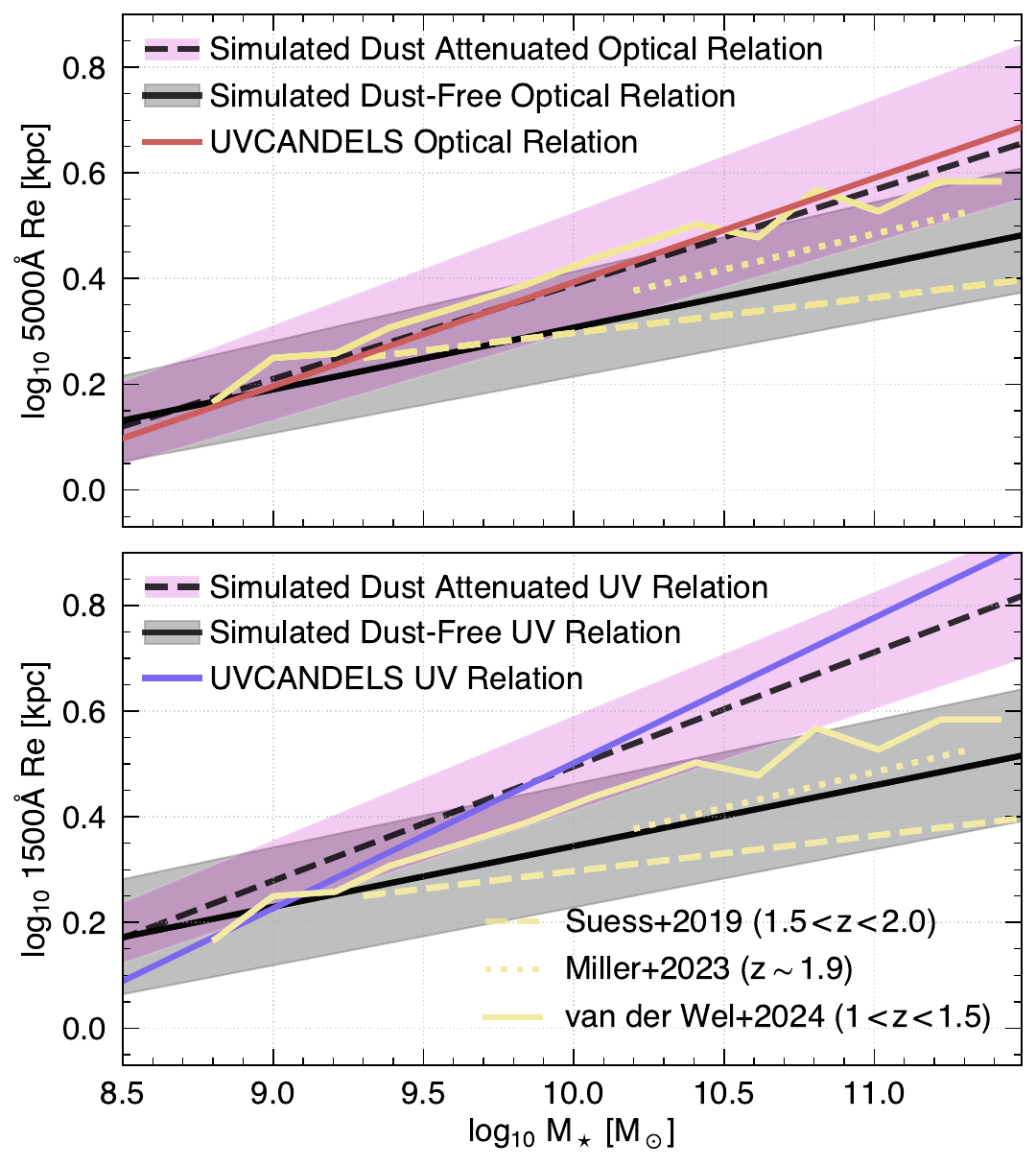}
    \vspace{-0.6cm}\caption{Stellar mass--size relations of mock VELA galaxies at $z=2$ reproduced from Figure~\ref{tab:mock_MRE} and compared to mass-weighted stellar mass--size relations from \protect\cite{Suess2019ApJ}, \protect\cite{Miller2023ApJ...945..155M}, and \protect\cite{vdwel2024ApJ}. The top and bottom panels show our relations in the optical and UV, respectively. }
    \label{fig:vela_MRe_mass_weighted}
\end{figure}

\begin{figure*}
    \centering
    \includegraphics[width=0.85\textwidth]{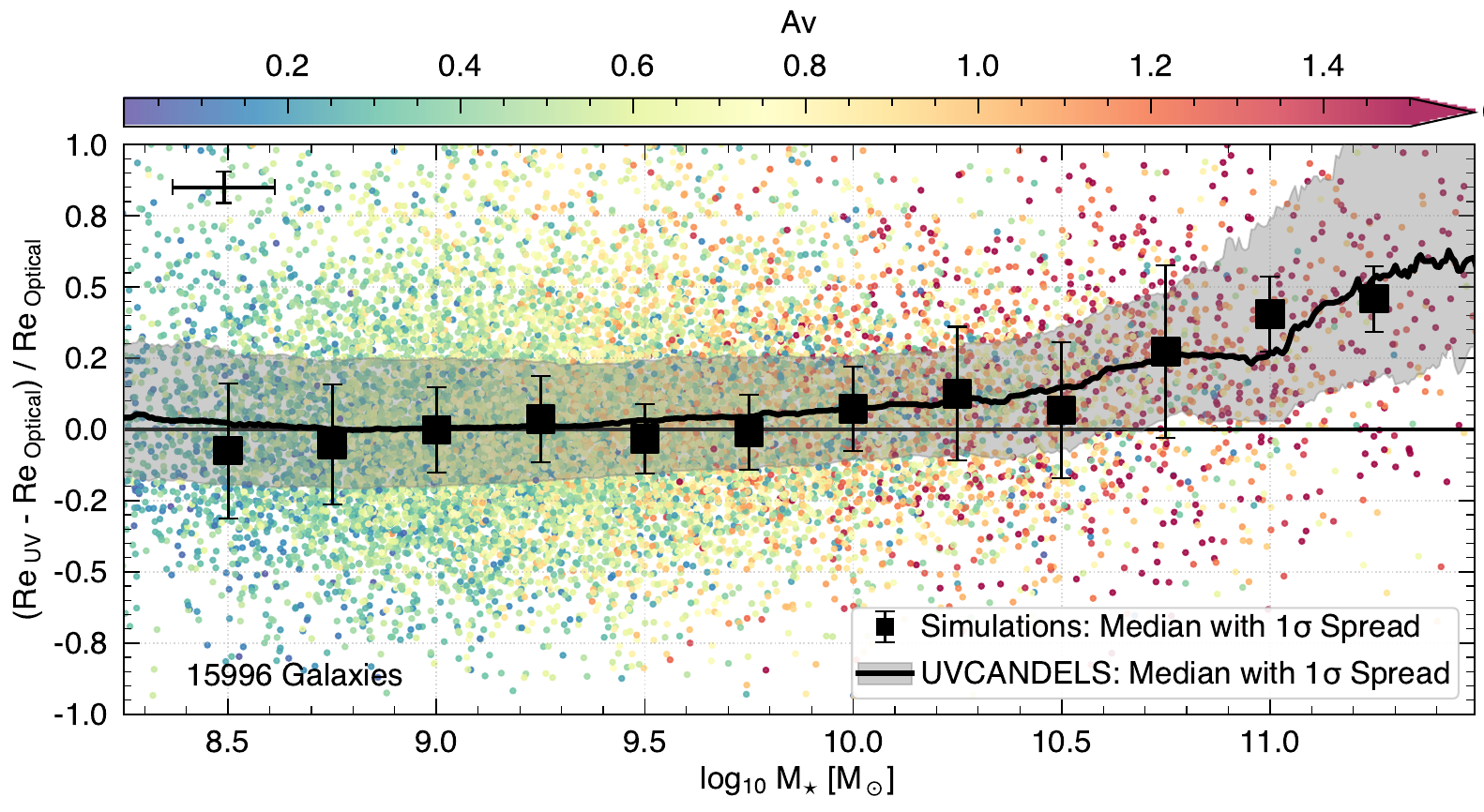}
    \caption{Comparison of rest-frame UV and optical sizes as in Figure~\ref{fig:size_diff} with the median trends with the simulated VELA galaxies at $z=2$ in Figure~\ref{fig:simulation_size_diff} shown as squares. As a function of stellar mass, we find remarkable agreement between the size differences measured from the simulations (squares) and UVCANDELS (black line). This agreement is evidence that differences between the sizes of massive galaxies in the UV and in the optical are primarily driven by dust attenuation.}
    \label{fig:simulation_comp}
\end{figure*}

If dust is primarily driving the color gradients in galaxies that are responsible for the differences between light- and mass-weighted sizes, then we may expect the dust-free stellar mass--size relation that we derive in Figure~\ref{fig:vela_MRe} to be consistent with mass-weighted stellar mass--size relations from the literature \citep[e.g.][]{Mosleh2017ApJ...837....2M, Suess2019ApJ, Suess2019ApJ...885L, Miller2023ApJ...945..155M, vdwel2024ApJ}. 
In Figure~\ref{fig:vela_MRe_mass_weighted}, we explore this by comparing the relations obtained for the mock galaxy sample to those derived for the UVCANDELS data, as well as several mass-weighted stellar mass--size relations. 
The relations in the rest-frame optical are shown in the top panel and in the rest-frame UV in the bottom panel. 
As in Figure~\ref{fig:vela_MRe}, the relations derived for the simulated sample where the effects of dust are included are shown as dashed lines with the magenta region encompassing all possible relations that have parameters with $1\sigma$ of the best-fit parameters listed in Table~\ref{tab:mock_MRE}.
Similarly, we show the same for the dust-free sample of mock galaxies as solid black lines and gray shaded regions.


We find that our best-fitting relations for the dust-free mock galaxies are in good agreement with the mass-weighted stellar mass--size relations from \cite{Suess2019ApJ}, \cite{Miller2023ApJ...945..155M}, and \cite{vdwel2024ApJ}, which are all shown in yellow in Figure~\ref{fig:vela_MRe_mass_weighted}. We note that most previous studies which suggest that mass-weighted radii provide a more unbiased probe of mass assembly in galaxies, have argued that light-weighted radii are strongly affected by color gradients. In this work, we show that once dust is accounted for, the stellar mass--size relation obtained by using half-light radii is in agreement with that when half-mass radii are used. This suggests that these color gradients are primarily driven by the effects of dust, consistent with \cite{Miller2023ApJ...945..155M}.
We also highlight that we still find the same trends in the stellar mass--size relation for the dust-free case, that is, that more massive galaxies tend to be larger in size. \cite{vdwel2024ApJ} similarly find that mass- and light-weighted sizes result in consistent pictures of size evolution over $0.5<z<2.3$ for both star-forming and quiescent galaxies using only observations.

In Figure~\ref{fig:simulation_comp}, we reproduce our results from Figure~\ref{fig:size_diff} using the UVCANDELS dataset and overplot the median and $1\sigma$ spread from the left panel of Figure~\ref{fig:simulation_size_diff}. This is done in order to showcase the remarkable agreement between the VELA simulations and UVCANDELS. Although there are some slight differences, particularly in the spread of the distribution at the high mass end, this difference is mostly caused by our low number of high mass galaxies in the VELA simulations.

At higher redshifts, which are now accessible thanks to JWST, \cite{Ormerod2024MNRAS.527.6110O} find that the stellar mass--size relation of galaxies breaks down at $z>3$ such that galaxy sizes vary little with mass. Although this relation may not hold at higher redshifts, \cite{Ono2024PASJ} show that the ratio between rest-frame UV and optical sizes is consistent with unity at $4\lesssim z \lesssim 10$ from Cosmic Evolution Early Release Science \citep[CEERS;][]{CEERS_Finkelstein_2023ApJ} imaging. In this paper, we show that if dust is removed from galaxies at $z=2$, the ratio between rest-frame UV and optical sizes should be consistent with unity as well. JWST is revealing many candidates with negligible dust content and extremely low dust attenuation at $z\gtrsim8$ \citep[e.g.][although see~\citealt{Rodighiero2023MNRAS} who also find heavily obscured galaxies at these redshifts]{Castellano2022ApJ, Finkelstein2022ApJ, Naidu2022ApJ, Atek2023MNRAS, Donnan2023MNRAS, Furtak2023MNRAS, Harikane2023ApJS, Tsuna2023MNRAS} possibly because dust is efficiently ejected during the early stages of galaxy formation \citep{Ferrara2023MNRAS}. Hence the findings from \cite{Ono2024PASJ} are
consistent with a picture in which color gradients found in galaxies at lower redshifts are primarily driven by dust. 
An alternative interpretation is that inside-out growth becomes a dominant evolutionary mechanism at $z\lesssim4$, such that these higher redshift galaxies do not have redder centers and bluer outskirts because they are evolving via a different mechanism. Although we show that at $z=2$ variations in galaxy size with wavelength do not constitute evidence for specific galaxy growth scenarios, extending these results to other redshifts requires additional results from simulations.

Using the same CEERS dataset, \cite{Suess2022ApJ} compare the sizes of galaxies at 1.5 $\mu$m and 4.4 $\mu$m, which roughly correspond to rest-frame optical and IR for galaxies at $1.0 \leq z \leq 2.5$. They find that galaxies are more compact at IR wavelengths and this size difference is much stronger for galaxies with stellar masses $\geq 10^{11}$ M$_\odot$. They demonstrate that spatial variations in age and attenuation are important, especially for more massive galaxies, consistent with the findings presented in this paper.

Currently, we do not have the necessary number statistics from simulations to correct our UVCANDELS measurements for the effects of dust. First, we are only able to make direct comparisons between rest-frame UV and optical sizes at $z=2$ as the bluest filter in which we have dust-free mock images is F435W. Dust-free mock images in bluer bands are necessary to extend this analysis to lower redshifts. Second, due to the relatively small number of galaxies in the VELA simulations, we find that any correction that we obtain is small and dominated by scatter due to variations between individual simulated galaxies. 

In order to properly correct galaxy sizes for the effects of dust at different wavelengths, inclination must also be considered as the effects of dust depend on the viewing angle. We have shown this explicitly for one example galaxy in Figure~\ref{fig:vela_dust_profiles}. Improved number statistics are therefore crucial as we only have a few galaxies at each orientation. Larger samples and mock imaging at bluer wavelengths become increasingly important if such a correction is to be applied as a function of redshift.

Finally, while the dust particles in the VELA simulation are distributed with a constant dust-to-metal ratio of 0.4 \citep{Simons2019ApJ}, recent results from \cite{Zhang2023MNRAS} indicate that high redshift galaxies likely have clumpy dust geometries and/or very compact dust cores. 
We posit that if these clumps and compact dust cores are not resolved, then invoking a modified dust geometry would likely leave the results presented in this paper largely unchanged. However, if they are resolved, then the slope of the rest-frame UV stellar mass--size relation from simulations would likely be even steeper than the optical relation. We have shown that centrally concentrated dust results in larger effective radii, especially in the rest-frame UV. Thus, very compact dust cores would flatten the measured UV Sérsic profile further, resulting in even larger rest-frame UV half-light radii for dusty galaxies. Additionally, the most massive clumps, which we would expect to also contain the most dust mass, have been shown to be located closer to the center of their host galaxy than less massive clumps \citep[e.g.][]{Martin2023ApJ} and therefore, both effects would likely result in more dust in the centers of galaxies and hence, larger rest-frame UV sizes than current simulations predict. Future work is needed to confirm the full effects of different dust geometries on galaxy sizes.







\section{Summary} \label{sec:conclusions}

In this paper, we have used UVCANDELS data to measure the rest-frame stellar mass -- size relation of disk galaxies at $0.5\leq z \leq3.0$. Rest-frame UV properties of galaxies at the lower end of this redshift range have remained largely unexplored due to a lack of deep high-resolution UV and blue-optical imaging. Thanks to the imaging obtained as part of UVCANDELS, such measurements are now possible and are presented in this paper. The addition these UV and B-band data are also essential for obtaining reliable measures of dust attenuation, which is a crucial component of this work. Our main conclusions are as follows. 
\begin{itemize}
\setlength{\itemsep}{-4pt}
    \item We present the UVCANDELS stellar mass--size relation in the rest-frame optical and UV in Figure~\ref{fig:mre}, finding that the relation in the UV is steeper across our entire redshift range. This difference is largest at $1.5\leq z\leq2.0$ and $2.0< z\leq2.5$, which are also the redshift bins in which we find the largest number of dust attenuated galaxies, as shown in Figure~\ref{fig:mre_av}.
    
    \item We find that the most massive galaxies are among the dustiest, as expected. These objects are also the ones that have significantly larger rest-frame UV sizes and are driving the difference in slope in the stellar mass--size relation between the rest-frame UV and optical.
    
    \item Using the VELA simulation suite, we investigate the stellar mass--size relations of synthetic dust-free galaxies and galaxies where the effects of dust are included. For the dust attenuated sample, we find remarkable agreement with our UVCANDELS results both in the rest-frame UV and optical. For the mock dust-free sample, we measure significantly shallower relations, directly showing that the presence of dust steepens the slope of stellar mass--size relation. This effect is stronger in the rest-frame UV, as shown in Figure~\ref{fig:vela_MRe} and Table~\ref{tab:mock_MRE}.

    \item We show that the shallower stellar mass--size relations for the dust-free mock sample are in general agreement with the mass-weighted stellar mass--size relations from \cite{Suess2019ApJ}, \cite{Miller2023ApJ...945..155M}, and \cite{vdwel2024ApJ} in Figure~\ref{fig:vela_MRe_mass_weighted}. This indicates that dust is largely responsible for the color gradients seen in disk galaxies.

    \item We show that once dust is accounted for, disk galaxies have consistent rest-frame UV and optical sizes (Fig.~\ref{fig:simulation_size_diff}). The presence of dust causes more massive galaxies to be larger in the rest-frame UV. In the VELA simulations, the UV size increase caused by dust closely matches our results from UVCANDELS, as shown in Figure~\ref{fig:simulation_comp}.

    \item Our results indicate that inside-out growth is not needed to explain the differences between rest-frame UV and optical size differences as the effects of dust alone can account for these differences. Future work will be needed to confirm this for larger simulated samples at different redshifts.
    
    \item Mock images that include and do not include the effects of dust attenuation for larger samples of galaxies over a wider wavelength range, and assuming different dust geometries, are needed to extend our analysis to other redshifts. These are also necessary to provide a reliable correction factor to the stellar mass--size relation of disk galaxies that accounts for the effects of dust. 
\end{itemize}

This paper has shown that, as a function of wavelength, the effects of dust on galaxy sizes and the stellar mass--size relation of disk galaxies are significant. However, plenty of scope for future work remains. As we have stressed throughout the paper, the simulations that we use -- despite being the best currently available for these purposes -- are limited both in wavelength (i.e.~the filters in which the mock observations are provided) and number statistics (i.e.~the number of galaxies in the simulation). With future advancements, understanding the effects of dust will be possible over large redshift ranges, now being explored with JWST, and for larger galaxy samples.

\begin{acknowledgments}
We thank the anonymous referee for thorough and constructive comments that have helped improve the quality and clarity of this paper. This work is based on observations with the NASA/ESA Hubble Space Telescope obtained at the Space Telescope Science Institute, which is operated by the Association of Universities for Research in Astronomy, Incorporated, under NASA contract NAS5- 26555. We acknowledge the generous support for Program numbers HST-GO-15647 and HST-AR-15798 that was provided through a grant from the STScI under NASA contract NAS5-26555. X.~W.~is supported by the Fundamental Research Funds for the Central Universities, the CAS Project for Young Scientists in Basic Research Grant No. YSBR-062, and the Xiaomi Young Talents Program.
\end{acknowledgments}

\vspace{5mm}
\facilities{HST(ACS, WFC3)}


\software{Astropy \citep{2013A&A...558A..33A,2018AJ....156..123A, Astropy3},  
          Source Extractor \citep{Bertin1996}
          }



\bibliography{my}{}
\bibliographystyle{aasjournal}



\end{document}